\documentclass[sigconf]{acmart}
\usepackage{amsmath,epsfig}

\long\def\comment#1{}

\usepackage[normalem]{ulem}

\usepackage{color}
\usepackage{graphicx}
\usepackage{xspace}
\usepackage{subfigure}
\usepackage{amsmath}
\usepackage{float}
\usepackage{graphicx}
\usepackage{url}
\usepackage{epsfig}
\usepackage{mathrsfs}
\usepackage{multirow}
\usepackage{color}
\usepackage{algorithmic}
\usepackage{listings}
\usepackage[nounderscore]{syntax}
\usepackage{setspace}
\usepackage{framed}
\usepackage{nopageno}
\usepackage{mathtools}
\usepackage[vlined,ruled,linesnumbered]{algorithm2e}
\usepackage{tikz}
\usepackage{fancybox}
\usepackage{eucal} 
\usepackage{pifont}

\usepackage{bm}  

\usepackage{balance}

\newcounter{remark}[section]
\renewcommand{\theremark}{\nthesection.\arabic{remark}}

\newcommand{\stitle}[1]{\vspace{1ex} \noindent{\bf #1}}

\newcommand{\kw}[1]{{\ensuremath {\mathsf{#1}}}\xspace}

\newcommand{\mat}[2]{{\begin{tabbing}\hspace{#1}\=\+\kill #2\end{tabbing}}}

\newcommand{\beqn}{\begin{eqnarray*}}
\newcommand{\eeqn}{\end{eqnarray*}}

\newcounter{ccc}
\newcommand{\bcc}{\setcounter{ccc}{1}\theccc.}
\newcommand{\icc}{\addtocounter{ccc}{1}\theccc.}

\newcommand{\kwnospace}[1]{{\ensuremath {\mathsf{#1}}}}

\newcommand{\SQL} {{\sl SQL}\xspace}

\newcommand{\PSM} {{\sl PSM}\xspace}
\newcommand{\rdbm} {{\sl RDBMS}\xspace}
\newcommand{\rdbms} {{\sl RDBMS}s\xspace}

\newcommand{\PostgreSQL} {{\sl PostgreSQL}\xspace}
\newcommand{\SQLServer} {{\sl SQL~Server}\xspace}

\newcommand{\Oracle} {{\sl Oracle}\xspace}

\newcommand{\WITH} {\kwnospace{with}\xspace}
\newcommand{\WITHplus} {\kwnospace{with}+\xspace}

\newcommand{\COMPUTEDBY} {\kw{computed}~\kw{by}}

\newcommand{\INSERT} {\kw{insert}}

\newcommand{\ALL} {\kw{all}}
\newcommand{\UNION} {\kw{union}}
\newcommand{\UUNION} {\kw{union}~\kw{by}~\kw{update}}

\newcommand{\PARTITIONBY} {\kw{partition}~\kw{by}}
\newcommand{\MAXREC} {\kw{maxrecursion}}

\newcommand{\SUM} {\kw{sum}}

\newcommand{\NORM} {\kw{norm}}

\newcommand{\Datalog} {{\sc Datalog}\xspace}
\newcommand{\XY} {{\sl XY}\xspace}

\newcommand{\TID}{I\!D\xspace}
\newcommand{\ID}{I\!D\xspace}

\newcommand{\uu}{\uplus}

\newcommand{\Spark}{{\sl Spark}\xspace}

\newcommand{\Hadoop}{{\sl Hadoop}\xspace}

\newcommand{\MAD}{{\sl MAD}\xspace}
\newcommand{\SimSQL}{{\sl SimSQL}\xspace}
\newcommand{\BISMARCK}{{\sl BISMARCK}\xspace}
\newcommand{\Morpheus}{{\sl Morpheus}\xspace}
\newcommand{\MauveDB}{{\sl MauveDB}\xspace}
\newcommand{\Psy}{{\sl Psycopg2}\xspace}
\newcommand{\SystemML}{{\sl SystemML}\xspace}
\newcommand{\BUDS}{{\sl BUDS}\xspace}

\newcommand{\SQLEM}{{\sl SQLEM}\xspace}
\newcommand{\USQL}{{\sl U-SQL}\xspace}

\newcommand{\UDF}{{\sl UDF}\xspace}
\newcommand{\UDFs}{{\sl UDF}s\xspace}
\newcommand{\UDA}{{\sl UDA}\xspace}

\newcommand{\GMM}{\kw{GMM}}
\newcommand{\MLR}{\kw{MLR}}
\newcommand{\MOE}{\kw{MOE}}

\setcopyright{none}

\begin{document}
\settopmatter{printacmref=false}
\title{Towards Expectation-Maximization by SQL in RDBMS}

\comment{
\author{%
{Kangfei Zhao, Jeffrey Xu Yu, Yu Rong}%
\vspace{1.4mm}\\
\fontsize{10}{10}\selectfont\itshape
The Chinese University of Hong Kong, Hong Kong SAR \\
\fontsize{9}{9}\selectfont\ttfamily\upshape
\{kfzhao, jiaosu, yu, hzhang\}@se.cuhk.edu.hk
}
}

\author{Kangfei Zhao}
\orcid{1234-5678-9012}
\affiliation{%
  \institution{The Chinese University of Hong Kong}
}
\email{kfzhao@se.cuhk.edu.hk}

\author{Jeffrey Xu Yu}
\affiliation{%
  \institution{The Chinese University of Hong Kong}
}
\email{yu@se.cuhk.edu.hk}

\author{Yu Rong}
\affiliation{%
  \institution{Tecent AI Lab}
}
\email{yu.rong@hotmail.com}

\author{Ming Liao}
\affiliation{%
  \institution{The Chinese University of Hong Kong}
}
\email{mliao@se.cuhk.edu.hk}

\author{Junzhou Huang}
\affiliation{%
  \institution{University of Texas at Arlington}
}
\email{jzhuang@uta.edu}

\def\thepage{\arabic{page}}
\pagestyle{plain}


\begin{abstract}
Integrating machine learning techniques into \rdbms is an important
task since there are many real applications that require modeling
(e.g., business intelligence, strategic analysis) as well as querying
data in \rdbms. Without integration, it needs to export the data from
\rdbms to build a model using specialized machine learning toolkits
and frameworks, and import the model trained back
to \rdbms for further querying. Such a process is not desirable since
it is time-consuming and needs to repeat when data is
changed. To support machine learning in \rdbms, there are proposals
that are platform-specific with limited functionalities to support
certain modeling.
In this paper, we provide an \SQL solution that has the potential to
support different machine learning modelings.  As an example, we study
how to support unsupervised probabilistic modeling, that has a wide
range of applications in clustering, density estimation and data
summarization, and focus on Expectation-Maximization (EM) algorithms,
which is a general technique for finding maximum likelihood
estimators.
To train a model by EM, it needs to update the model parameters by an
E-step and an M-step in a while-loop iteratively until it converges to
a level controled by some threshold or repeats a certain number of
iterations.  To support EM in \rdbms, we show our answers to the
matrix/vectors representations in \rdbms, the relational algebra
operations to support the linear algebra operations required by EM,
parameters update by relational algebra, and the support of a
while-loop. It is important to note that the \SQL'99 recursion cannot
be used to handle such a while-loop since the M-step is non-monotonic.
In addition, assume that a model has been trained by an EM algorithm,
we further design an automatic in-database model maintenance mechanism
to maintain the model when the underlying training data changes. We
have conducted experimental studies and will report our findings in
this paper.
\end{abstract}

\maketitle
\thispagestyle{plain}

\section{Introduction}
Nowadays, integrating advanced data analytical techniques into \rdbms
is an urgent requirement for data
integration~\cite{DBLP:conf/sigmod/DongHMN05}, business intelligence
and strategic analysis~\cite{DBLP:journals/misq/ChenCS12,
  DBLP:journals/corr/MoniruzzamanH13}. Among these techniques that
need to be integrated into \rdbms, machine learning models play a
leading role in predictive and estimation tasks.  Although many
specialized machine learning toolkits and frameworks (e.g.,
scikit~\cite{sklearn} and
TensorFlow~\cite{DBLP:conf/osdi/AbadiBCCDDDGIIK16}) are designed and
developed, the approaches of building, utilizing and managing machine
learning models in \rdbms still need a comprehensive exploration.
First, in most enterprise applications, data are stored in a database
system. It is cumbersome and time-consuming of exporting the data from
the database system and then feeding it into models, as well as
importing the prediction and estimation results back to the database system.
Second, it is highly desirable that users can build a model
as to query data in \rdbms, and query their data by exploiting the
analysis result of the models trained as a part of query in a seamless
similar in \rdbms. What we need is a flexible way to train/query a
machine learning model together with data querying by a high-level
query language (e.g., \SQL).
Third, the data maintained in \rdbms is supposed to change, and there
is more data collected from time to time frequently. The analysis
result of the models trained in a machine learning toolkit/framework
may be out-dated, which requires to repeat the process of exporting
data from \rdbms followed by importing the model trained into \rdbms.
Given the fact that \rdbms have the techniques (e.g., trigger) to
manage data updating automatically when data changes, a further
consideration is how to manage the machine learning models update
automatically using the database techniques available.

There are efforts to support machine learning in \rdbms.  Early
in-database machine learning is developed based on {\UDF}s or
specific libraries like
MADlib~\cite{DBLP:journals/pvldb/HellersteinRSWFGNWFLK12} for
\PostgreSQL, Oracle Data
Mining~\cite{DBLP:books/sp/datamining2005/TamayoBCYMMTHKTKMHSM05}, DB2
Intelligent Miner, etc.  On one hand, these functions and libraries
can achieve optimized performance. On the other hand, they are
platform-specific and have limited functionalities from the high-level
syntax to the low-level implementation.  It is difficult for database
end-users to extend these libraries to support their own models that are
not available in the libraries.  To fulfill logical and physical data
isolation, model-based views~\cite{DBLP:conf/sigmod/DeshpandeM06,
  DBLP:journals/pvldb/KocR11} are proposed to support classification
and regression analysis in database systems.  Like regular views,
model-based views support querying, materialization, and maintenance
strategies.  In brief, \cite{DBLP:conf/sigmod/DeshpandeM06,
  DBLP:journals/pvldb/KocR11} allow using an ad-hoc \kw{create~view}
statement to declare a classification view.  In this \kw{create~view}
statement, \cite{DBLP:conf/sigmod/DeshpandeM06} specifies the model by
an \kw{as ... fit... bases} clause, and the training data is fed by an
\SQL query, while \cite{DBLP:journals/pvldb/KocR11} specifies a model
explicitly with \kw{using~svm} clause, where the features and labels
are fed by \kw{feature~function} and \kw{labels}, respectively. Here,
\kw{feature~function} takes database attributes as the input features,
and labels are database attributes.
Although these approaches provide optimized implementation for
classification models, their \kw{create~view} statement is lack of
generality and deviating from the regular \SQL syntax.  In addition,
the models supported are limited and implemented in a low-level form
in a database system, which makes it difficult for ordinary database
end-users to develop new models swiftly.
In this work,  
we demonstrate our SQL recursive query can define a model-based view
in an explicit fashion and can be used to support many machine
learning models. Different from \cite{DBLP:conf/sigmod/DeshpandeM06,
  DBLP:journals/pvldb/KocR11}, we focus on unsupervised models in the
application of in-database clustering, density estimation and data
summarization.

We take a two-step approach to support machine learning in \rdbms. The
first step is to design in-database machine learning framework based
on \SQL, in particular \SQL recursive queries, to show that \SQL has
its capability of doing machine learning by \SQL in \rdbms.  The
second step is to further find an efficient way to support queries for
machine learning in \rdbms. In this paper, we concentrate on
the first step, following our previous work to support graph analytics
in \rdbms using \SQL recursive queries~\cite{conf/sigmod/ZhaoY17}. The
focus of this paper is on how to train machine learning models in
\rdbms, given that the core computations of model training are
linear algebra operations and parameter updating.

Consider training a machine learning model. In brief, it has an
initial parameter setting for the model, and will update the
parameters in a while-loop iteratively until it converges to a level
controled by some threshold or repeats a certain number of
iterations. The model trained is the model with the parameters
obtained at the end of the while-loop.  To do so in \rdbms, there are
several things that need to be handled: the ways to represent
matrix/vector in \rdbms, the relational algebra operations to support
the linear algebra operations required, the way to update parameters,
and the support of a while-loop. In this paper, we provide our answer
to such needs.

The main contributions of this work are summarized below.
First, we study in \rdbms how to support unsupervised probabilistic
modeling, that has a wide range of applications in clustering, density
estimation and data summarization, and focus on
Expectation-Maximization (EM)
algorithms~\cite{mclachlan2007algorithm}, which is a general technique
for finding maximum likelihood estimators. In EM, the parameters to be
trained are means, covariances, and mixing coefficients; there
are two main steps in a while-loop, namely, E-step for expectation and
M-step for maximization, and the parameters are updated in the
while-loop.
Second, we discuss how to represent data in \rdbms in different ways,
how to compute E-step and M-step using relational algebra operations
(e.g., natural join, group-by and aggregation), how to update
parameters using relational algebra operations, and how to support the
while-loop using \SQL recursive queries.
It is worth mentioning that the recursion for EM is a mutual recursion
of E-step and M-step. Recall that the E-step is to compute the
conditional posterior probability by Bayesian inference, which can be
supported by \SQL as a monotonic operation, whereas the M-step is to
compute and update the parameters of the model given a closed-form
updating formula, which cannot be monotonic. This fact suggests that
\SQL'99 recursion cannot be used to support EM, since \SQL'99
recursion (e.g., recursive \WITH) only supports stratified negation,
and therefore cannot support non-monotonic operations. We use
\XY-stratified~\cite{ZanioloAO93,zaniolo1997advanced,ArniOTWZ03}, and
provide an enhanced \SQL recursion (e.g., \WITHplus), which can handle
non-monotonic operations. We have implemented our approach as an \SQL
layer on top of \PostgreSQL, and process our \WITHplus using
\PostgreSQL. We show how to train a batch of classical statistical
models~\cite{DBLP:books/lib/Bishop07}, including Gaussian Mixture
model, Bernoulli Mixture model, mixture of linear regression, Hidden
Markov model, Mixtures of Experts, using the recursive \SQL queries.
\comment{
implemented by \SQL analytic functions, which is a monotonic operation
in the recursive query. The M-step, computing and updating the
parameters of the model given a closed-form updating formula, are
supported by the MV/MM-join, union-by-update, and the 6 basic
relational algebra operations. According to
\cite{conf/sigmod/ZhaoY17}, any closed-form updating formula which can
be supported by above operations, plus the monotonic \SQL analytic
functions in Bayesian inference, make sure the mutual recursive query
has an iterative fixed point.
}
%
Third, Given a model trained by an EM algorithm, we further design an
automatic in-database model maintenance mechanism to maintain the
model when the underlying training data changes. Inspired by the
online and incremental EM algorithms~\cite{DBLP:conf/naacl/LiangK09,
  neal1998view}, we show how to obtain the sufficient statistics of
the models to achieve the incremental even decremental model updating,
without re-building the model using all data. It is worth mentioning
that our setting is different. Different from the incremental EM
algorithms which are designed to train the model during its iterative
processing, we re-train the model by sufficient statistics using
partial data being used to build the previous model in addition to the
new data.
Fourth, we have conducted experimental studies and will report our
findings in this paper.

\stitle{Organization} Section \ref{sec:rw} discusses the related
works. In Section~\ref{sec:rq}, we introduce the preliminaries
including the EM algorithm and the requirements to support it in
database systems. Then, our solution is sketched in
Section~\ref{sec:oursolu} and the \SQL recursive query and the
implementation details are introduced in
Section~\ref{sec:recforem}. In Section~\ref{sec:maintain}, we design a
view update mechanism, which is facilitated by triggers. We conduct
extensive experimental studies in Section~\ref{sec:exp} and conclude
the paper in Section~\ref{sec:conclusion}.


\section{Related Works}  
\label{sec:rw}
Our study is closely related to the following research topics:
 
\stitle{Machine Learning with \SQL}. 
There are a board discussion on the approaches to using \SQL for ML, which are at different levels of abstraction in the long-term research. In early years,  Ordonez et al. presents pure \SQL implementation of the EM algorithm in \rdbm, including K-means~\cite{DBLP:journals/tkde/Ordonez06} and Gaussian Mixture Model~\cite{DBLP:conf/sigmod/OrdonezC00}.
Their approach, \SQLEM~\cite{DBLP:conf/sigmod/OrdonezC00}, presents three strategies to implement EM in \SQL: horizontal, vertical and a hybrid one. However these implementations cannot support high dimensional data and a large cluster number effectively.
Computations involving matrix and vector primitives are expressible in \SQL with the aid of \UDFs. For example, \MAD \cite{DBLP:journals/pvldb/CohenDDHW09, DBLP:journals/pvldb/HellersteinRSWFGNWFLK12} is a in-database analytics library for matrix and vector operators. 
Luo et. al.~\cite{DBLP:conf/icde/LuoGGPJ17} extend \SimSQL~\cite{DBLP:conf/sigmod/CaiVPAHJ13}, a Hadoop-based relational database system to enable linear algebra computations.
Taking the functions manipulating matrix/vector data type as a set of building blocks, a \SQL query can support basic machine learning task, e.g., least square regression.
Furthermore, \UDF and \UDA can be used to implement gradient methods~\cite{DBLP:journals/debu/BorkarBCRPCWR12, DBLP:conf/sigmod/FengKRR12}. \MAD and \BISMARCK~\cite{DBLP:conf/sigmod/FengKRR12} use python \UDF and \UDA to support stochastic gradient decent, respectively. 
To deploy applications of stochastic models and analytics techniques, the monte carlo database system~(MCDB)~\cite{DBLP:journals/tods/JampaniXWPJH11, DBLP:conf/sigmod/CaiVPAHJ13} provides stochastic models to be directly used with the data stored in a large database.In MCDB, a \UDF called value-generating (VG) function is used to draws samples from databases and bayesian learning can be performed by \SQL queries subsequently. 

\stitle{Query Optimization for Machine Learning}. 
Using database techniques to improve the efficiency of machine learning
application is a research focus currently. These techniques mainly aim at minimizing the computational redundancy incurred by 
the extra storage for normalized data and sparse data. 
\cite{DBLP:conf/sigmod/SchleichOC16, DBLP:conf/sigmod/KumarNP15} learn linear models over multi-table normalized data, introducing the idea of learning on factorized database. 
\Morpheus~\cite{DBLP:journals/pvldb/ChenKNP17}, a database middleware, can automatically convert the linear algebra operators of denormalized matrix/vector to normalized data by a set of rewriting rules. 
\cite{DBLP:conf/sigmod/LiCZ00NP19} utilizes tuple-oriented compression to reduct the data redundancy for mini-batch stochastic gradient descent.  
In addition, in the ML system design,  logical and physical plan optimization techniques, like plan simplification and rewriting~\cite{DBLP:conf/pods/KhamisNR16, DBLP:journals/pvldb/BoehmDEEMPRRSST16, DBLP:conf/edbt/KernertKL15}, operator selection~\cite{DBLP:journals/pvldb/BoehmDEEMPRRSST16}, physical operator fusion~\cite{DBLP:journals/pvldb/BoehmDEEMPRRSST16, DBLP:conf/cidr/ElgamalLBETRS17}, delta updating rules~\cite{DBLP:conf/sigmod/NikolicEK14} are widely used to improve the performance of the system. 
These optimization techniques for linear algebra and matrix calculus provides a large potential to improve the performance of model-based view in different scenarios. 

\stitle{Declarative Language for Machine Learning}.
Apart from pure \SQL, there have been some efforts aimed at building statistical and machine learning applications by a declarative language in database and data processing systems. 
Microsoft Azure Data Lake Analytics provides an extension of \SQL, named \USQL~\cite{klein2017u} with a tight C\# binding to support distributed machine learning. 
Similarly, \SystemML~\cite{DBLP:journals/pvldb/BoehmDEEMPRRSST16} expresses machine learning algorithms by a simplified R and python like-language, and automatically translates the program into execution plan on top of \Spark. 
\BUDS~\cite{DBLP:conf/sigmod/GaoLPJ17} is a language for Bayesian machine learning, specifically, Markov chain simulation, allowing distributed computation on types of sets, maps, vectors and matrices.
In addition, \Datalog and its extensions~\cite{DBLP:journals/pvldb/LiCCWZ17, eisner-2008, DBLP:journals/debu/BorkarBCRPCWR12} are also used to integrate statistical and machine learning into data management systems.
In this paper, we focus on \SQL query, as it is the most widely-used declarative query language in database. We show that users can build model-based view by \SQL recursive query with limited enhancement and the support of vector/matrix data type.

\stitle{Model-based View in \rdbm}. 
To provide adequate support for modeling data in database system, the abstraction of model-based view is proposed to persist the model in the database and data mining area. 
\MauveDB~\cite{DBLP:conf/sigmod/DeshpandeM06} is an architecture which supports a SQL-based declarative language to define  views for regression model and interpolation. 
Koc et. al.~\cite{DBLP:journals/pvldb/KocR11} define and maintain statistical model for classification in \rdbms, including least square regression, logistic regression, ridge regression and SVM. 
Nikolic et. al.~\cite{DBLP:conf/sigmod/NikolicO18} propose a unified incremental view maintenance approach for factorized database, which can support liner regression model with gradient methods. Most of the existing studies focus on the supervised statistical model while few attention is paid on supporting and maintaining unsupervised modeling in database system. 

\section{Preliminaries}
\label{sec:rq}

In this paper, we focus on unsupervised probabilistic modeling, which has broad applications in clustering, density estimation and data summarization in database and data mining area. Specifically, the unsupervised models aim to reveal the relationship between the observed data and some latent variables by maximizing the data likelihood. The expectation-maximization~(EM) algorithm, first introduced in \cite{dempster1977maximum}, is a general technique for finding maximum likelihood estimators. 
It has a solid statistical basis, robust to noisy data and its complexity is linear in data size. 
Here, we use the Gaussian mixture model~\cite{DBLP:books/lib/Bishop07}, a widely used model in data mining, pattern recognition, and machine learning, as an example to illustrate the EM algorithm and our approach throughout this paper. 

\begin{algorithm}[t]
\caption{{EM Algorithm for Mixture Gaussian Model}}
\label{algo:em}
\begin{algorithmic}[1]
\STATE  Initialize the means $\bm{\mu}$, covariances $\bm{\sigma}$ and mixing coefficients $\bm{\pi}$;
\STATE  Compute the initial log-likelihood $L$, $i \leftarrow 0$;
\WHILE {$\Delta L > \epsilon $ \OR $i <$ \MAXREC}
	\STATE {E-step: compute the responsibilities $p(z_{ik})$ based on current $\bm{\mu}$, $\bm{\sigma}$ and $\bm{\pi}$ by Eq.~(\ref{gmm:estep});}
	\STATE {M-step: re-estimate $\bm{\mu}$, $\bm{\sigma}$ and $\bm{\pi}$ by Eq.~(\ref{gmm:means})-(\ref{gmm:mixingcoef});}
	\STATE {re-compute the log-likelihood $L$; $i \leftarrow i + 1$;}
\ENDWHILE
\RETURN $\bm{\mu}$, $\bm{\sigma}$, $\bm{\pi}$;
\end{algorithmic}
\end{algorithm}

Suppose we have an observed dataset $\bm{X} = \{ x_1, x_2, \cdots, x_n\}$ of n data points where $x_{i} \in \mathbb{R}^{d}$. 
Given $\mathcal{N}(x | \bm{\mu}, \bm{\sigma})$ is the probability density function of a Gaussian distribution with mean $\bm{\mu} \in \mathbb{R}^{d}$ and covariance $\bm{\sigma} \in \mathbb{R}^{d \times d}$, the density of Gaussian mixture model is a simple linear super-position of $K$ different Gaussian components in the form of 
Eq.~(\ref{eq:gmm}).
\begin{equation}
p(x_{i}) = \sum_{k = 1}^{K} \pi_k \mathcal{N}(x_{i} | \bm{\mu}_{k}, \bm{\sigma}_{k})
\label{eq:gmm}
\end{equation}
Here, $\pi_{k} \in \mathbb{R}$ is the mixing coefficient, i.e., the prior of a data point belonging to component $k$ and satisfies $\sum_{i = 1}^{K} \pi_{k} = 1$. To model this dataset $X$ using a mixture of Gaussians, the objective is to maximize the log of the likelihood function in Eq.~(\ref{eq:loglikelihood}).
\begin{equation}
ln p(\bm{X} | \bm{\pi}, \bm{\mu}, \bm{\sigma}) = \sum_{i = 1}^{n} ln [ \sum_{k = 1}^{K} \pi_k \mathcal{N}(x_i | \bm{\mu}_{k}, \bm{\sigma}_{k}) ]
\label{eq:loglikelihood}
\end{equation}
Algorithm~\ref{algo:em} sketches the EM algorithm for training the Gaussian Mixture Model.
First, in line 1-2, the means $\bm{\mu}_k$, covariances $\bm{\sigma}_k$ and the mixing coefficients $\bm{\pi}_k$ of $K$ Gaussian distributions are initialized, and the initial value of the log-likelihood (Eq.~(\ref{eq:loglikelihood})) is computed. 
In the while loop of line 3-7, the Expectation-step (E-step) and Maximization-step (M-step) are executed alternatively. 
In the E-step, we compute the responsibilities, i.e., the conditional probability that $x_{i}$ belongs to component $k$, denoted as $p(z_{ik})$ by fixing the parameters based on the Bayes rule in Eq.~(\ref{gmm:estep}).
\begin{equation}
p(z_{ik}) = \frac{\pi_k \mathcal{N}(x_i | \bm{\mu}_k, \bm{\sigma}_k)}
{\sum_{j = 1}^{K} \pi_j \mathcal{N}(x_i | \bm{\mu}_j, \bm{\sigma}_j)}
\label{gmm:estep}
\end{equation}
In M-step, we re-estimate a new set of parameters using the current responsibilities by maximizing the log-likelihood (Eq.~(\ref{eq:loglikelihood})) as follows. 
\begin{align}
\label{gmm:means} \bm{\mu}_k^{new} & = \frac{1}{n_k} \sum_{i = 1}^{n} p(z_{ik}) x_i  \\
\label{gmm:covs} \bm{\sigma}_k^{new} & = \frac{1}{n_k} \sum_{i = 1}^{n} p(z_{ik}) (x_i - \bm{\mu}_k^{new})  (x_i - \bm{\mu}_k^{new})^T  \\
\label{gmm:mixingcoef} \bm{\pi}_k^{new} & = \frac{n_k}{n}
\end{align}
where $n_k = \sum_{i = 1}^{n} p(z_{ik})$.
At the end of each iteration, the new value of log-likelihood is evaluated and used for checking convergence. The algorithm ends when the log-likelihood converges or a given iteration time is reached.
In \rdbm, the learnt model, the parameters of $K$ components, can be persisted in a relation of $K$ rows as shown in Table~\ref{tbl:rels}(a). Suppose 1-dimensional dataset $X$ as Table~\ref{tbl:rels}(b) is given, the posterior probability of $x_{i}$ belongs to component $k$ can be computed as Table~\ref{tbl:rels}(c) and clustering can be conducted by assigning $x_{i}$ to component with the maximum $p(z_{ik})$.
 
\begin{table}[t]
\begin{footnotesize}
\begin{minipage}[t]{2.4cm}
\begin{center}
\begin{tabular}{|l|l|l|l|} \hline 
$K$ & $\bm{\pi}$ & $\bm{\mu}$ & $\bm{\sigma}$ \\ \hline \hline
1 & $\pi_{1}$ & $\mu_{1}$ & $\sigma_{1}$ \\ \hline
2 & $\pi_{2}$ & $\mu_{2}$ & $\sigma_{2}$ \\ \hline
\end{tabular} \\
\vspace*{0.2cm}
(a) Relation GMM
\vspace*{0.2cm}
\end{center}
\end{minipage}
\quad
\begin{minipage}[t]{2cm}
\begin{center}
\begin{tabular}{|c|c|} \hline 
$\TID$ & $x$ \\ \hline \hline
1 & $x_1$ \\ \hline
2 & $x_2$ \\ \hline
\end{tabular} \\
\vspace*{0.2cm}
(b) Relation $X$
\vspace*{0.2cm}
\end{center}
\end{minipage}
\quad
\begin{minipage}[t]{2.4cm}
\begin{center}
\begin{tabular}{|l|l|l|} \hline 
$\TID$ & $K$ & $p$ \\ \hline \hline
1 & 1 & $p(z_{11})$ \\ \hline
1 & 2 & $p(z_{12})$ \\ \hline
2 & 1 & $p(z_{21})$ \\ \hline
2 & 2 & $p(z_{22})$ \\ \hline
\end{tabular} \\
\vspace*{0.2cm}
(c) Relation $R$
\vspace*{0.2cm}
\end{center}
\end{minipage}
\centering
\vspace*{0ex}
\caption{The relation representations}
\label{tbl:rels}
\end{footnotesize}
\end{table}

To fulfil the EM algorithm in database systems, there are several important issues need to be concerned, including  
(1) the representation and storage of high dimensional data in database.  
(2) the relation algebra operation used to perform linear algebra computation in EM.
(3) the approach for iterative parameter updating.
(4) the way to express and control the iteration of EM algorithm.
(5) the mechanism to maintain the existing model when underlying data involves.  

As an early attempt, Ordonez et. al.~\cite{DBLP:conf/sigmod/OrdonezC00} proposed a \SQL implementation, \SQLEM. 
Their implementation is based on two strategies: horizontal and vertical. 
The horizontal approach organizes the data points as the horizontal representation as shown in Table~\ref{tbl:represent}(b), where the relation has $n$ rows and $d$ attributes. 
The vertical approach organizes the data points as the coordinate representation as shown in Table~\ref{tbl:represent}(a), where the relation has $nd$ rows, 3 attributes for the identity of data, index and value.
The horizontal approach has efficient performance while the vertical approach has flexible expression power. Thereby, \cite{DBLP:conf/sigmod/OrdonezC00} further proposes a hybrid approach, which persists both vertical and horizontal representations simultaneously to achieve a tradeoff. 
Although it provides a feasible solution for EM algorithm in \rdbm, their approach has some drawbacks.
First, the hybrid and vertical approaches only have limited flexibility in expressing linear algebra operations. 
For the hybrid approach, the users still need to specify the computation for each dimension $d$ in the E-step and for each $k$ in the M-step, leading to many repetitive and complicated expressions in the \SQL queries. These queries also limit the supported models. For example, the covariance matrix $\bm{\sigma}$ must be diagonal.
Second, it has not provided an effective parameter updating and iteration control mechanism inside the database system. The \SQL queries for one iteration E-step and M-step are wrapped in a while-loop of a host language, i.e., Java, python. This will undermine the overall performance due to large communication and I/O cost. 
Furthermore, their approach does not support model updating. 

\comment{
To support the M-step of Eq.~(\ref{gmm:means}) and (\ref{gmm:covs}), the algebraic structure, namely semiring,
is shown to have sufficient expressive power to support matrix/vector computation.
%
%
The semiring is a set of ${\mathcal M}$ including two identity
elements, ${\bf 0}$ and ${\bf 1}$, with two operations: addition ($+$)
and multiplication ($\cdot$). In brief, (1) $({\mathcal M}, +)$ is a
commutative monoid with ${\bf 0}$, (2) $({\mathcal M}, \cdot)$ is a monoid
with ${\bf 1}$, (3) the multiplication ($\cdot$) is left/right
distributes over the addition ($+$), and (4) the multiplication by
${\bf 0}$ annihilates ${\mathcal M}$.
Below, $\kw{A}$ and $\kw{B}$ are two $2 \times2$ matrix,
and $\kw{C}$ is a vector with $2$ elements.
 
\begin{eqnarray*}
  \kw{A} \cdot \kw{B}
  &=&
{\small
  \left(
\begin{array}{cc}
  a_{11} \odot b_{11} \oplus a_{12} \odot b_{12} \\
  a_{21} \odot b_{11} \oplus a_{22} \odot b_{22}
\end{array}
\right)
}
\\
  \kw{A} + \kw{B}
  &=&
{\small
  \left(
\begin{array}{cc}
  a_{11} \oplus b_{11} & 
  a_{12} \oplus b_{12} \\
  a_{21} \oplus b_{21} & 
  a_{22} \oplus b_{22} 
\end{array}
\right)
}
\\
\kw{A} \cdot \kw{C}
&=&
\left(
\begin{array}{c}
  a_{11} \odot c_1 \oplus a_{12} \odot c_2  \\
  a_{21} \odot c_1 \oplus a_{22} \odot c_2 
\end{array}
\right)
\end{eqnarray*}
We focus on the multiplication ($\cdot$), since it is trivial to
support the addition ($+$) in relational algebra.
Let $\kw{A}$ and $\kw{B}$ be two $n \times n$ matrices, and $\kw{C}$ be a
vector with $n$ elements. For the multiplication $\kw{AB} = \kw{A}
\cdot \kw{B}$, and $\kw{AC} = \kw{A} \cdot \kw{C}$, we have the
following.

\vspace*{-0.0cm}
\begin{small}
\begin{eqnarray}
  \kw{AB}_{i} &=& \mathop{\bigoplus}_{j=1}^{n} \kw{A}_{ij} \odot \kw{B}_{ij}
  \label{eq:mm0}
  \\
  \kw{AC}_{i} &=& \mathop{\bigoplus}_{j=1}^{n} \kw{A}_{ij} \odot \kw{C}_{j}
  \label{eq:mv0}
\end{eqnarray}
\end{small}
\vspace*{-0.0cm}
}

\section{Our Solution}
\label{sec:oursolu}
In this section, we propose a complete solution to deal with above issues in applying the EM algorithm and building model-based views inside \rdbm.  
\comment{
A model-based view is a materialized view which persists the parameters, $\bm{\mu}$, $\bm{\sigma}$ and $\bm{\pi}$ estimated by the iterative E-step and M-step above. The view trained from a given dataset plays a same role as conventional database view, and can be maintained with incremental changes of the training set.

To support the model-based view, the database system should internally support (1) statistical functions (2) matrix/vector data type and calculus (3) iterative/recursive evaluation.
}

\begin{table}[t]
\begin{footnotesize}
\begin{minipage}[t]{2.4cm}
\begin{center}
\begin{tabular}{|l|l|l|} \hline 
$\TID$ & $C$ & $x$ \\ \hline \hline
1 & 1 & $1.0$ \\ \hline
1 & 2 & $2.0$ \\ \hline
2 & 1 & $3.0$ \\ \hline
2 & 2 & $4.0$ \\ \hline
\end{tabular} \\
\vspace*{0.2cm}
(a) coordinate representation $X_a$
\vspace*{0.2cm}
\end{center}
\end{minipage}
\quad
\begin{minipage}[t]{2.4cm}
\begin{center}
\begin{tabular}{|l|l|l|} \hline 
$\TID$ & $d_{1}$ & $d_{2}$ \\ \hline \hline
1 & 1.0 & $2.0$ \\ \hline
2 & 3.0 & $4.0$ \\ \hline
\end{tabular} \\
\vspace*{0.2cm}
(b) horizontal representation $X_b$
\vspace*{0.2cm}
\end{center}
\end{minipage}
\quad
\begin{minipage}[t]{2.4cm}
\begin{center}
\begin{tabular}{|l|l|} \hline 
$\TID$ & $x$ \\ \hline \hline
1 & $[1.0, 2.0]$  \\ \hline
2 & $[3.0, 4.0]$  \\ \hline
\end{tabular} \\
\vspace*{0.2cm}
(c) row-major representation $X_c$
\vspace*{0.2cm}
\end{center}
\end{minipage}
\centering
\vspace*{0ex}
\caption{The relation representations for matrix}
\label{tbl:represent}
\end{footnotesize}
\end{table}

\stitle{High Dimensional Representation}. 
Regarding the issue of high dimensional data, different from ~\cite{DBLP:conf/sigmod/OrdonezC00}, we adopt the row-major representation, as shown in Table~\ref{tbl:represent}(c), which is endorsed by allowing array/vector data type in database. 
For one thing, this keeps the efficient performance of horizontal representation by reducing I/O cost, especially for dense vectors.  
For the other, we can use the vector/matrix operations to support complicated linear algebra computation in a concise \SQL query. 
As most \rdbms have provided the array/vector datatype internally, apart from the build-in array functions, 
many extended libraries of database~\cite{DBLP:journals/pvldb/CohenDDHW09, DBLP:books/sp/datamining2005/TamayoBCYMMTHKTKMHSM05} also provide additional statistical function and vector/matrix operations for multivariable statistical analysis and basic linear algebra calculus. These high-level abstractions avoid letting end-users specify the arithmetic operations on each dimension of the data point so that serve as a set of building blocks of machine learning algorithms.

Consider computing the means $\bm{\mu}$ in the M-step (Eq.~(\ref{gmm:means})) with the 3 different representations $X_a$, $X_{b}$ and $X_{c}$ in Table~\ref{tbl:represent}. Suppose the responsibilities are in relation $R(\TID, K, p)$, where $\TID$, $K$ and $p$ is the identifier of data point and component, and the value of $p(z_{ik})$.
The relational algebra expressions to compute Eq.~(\ref{gmm:means}) are shown in Eq.~(\ref{eq:gmm:means:coo})-(\ref{eq:gmm:means:row}), respectively. 
\begin{align}
\displaystyle{ M_a } & \leftarrow  \rho_{(K, C, \text{mean})} ({}_{K, C}{\mathcal{G}_{\SUM(p * x)}}(R \mathop{\Join}_{\substack{R.\TID = X_a.\TID}} X_a))  \label{eq:gmm:means:coo} \\
\displaystyle{ M_b } & \leftarrow  \rho_{(K, d1, d2)} ({}_{K}{\mathcal{G}_{\SUM(p * d1), \SUM(p * d2)}}(R \mathop{\Join}_{\substack{R.\TID = X_b.\TID}} X_b)  \label{eq:gmm:means:hor}) \\
\displaystyle{ M_c } & \leftarrow  \rho_{(K, \text{mean})} ({}_{K}{\mathcal{G}_{\SUM(p \cdot x)}}(R \mathop{\Join}_{\substack{R.\TID = X_c.\TID}} X_c))  \label{eq:gmm:means:row}
\end{align} 
We elaborate on these expressions. In Eq.~(\ref{eq:gmm:means:coo}) and Eq.~(\ref{eq:gmm:means:hor}), the binary operator $*$ is the arithmetic multiplication while in Eq.~(\ref{eq:gmm:means:row}), the operator $\cdot$ denotes a scalar-vector multiplication. First, all of these 3 representations need to join $X$ and $R$ on the $\TID$ attribute to compute $p(z_{ik})x_i$. 
The differences lay in the group and aggregation for each component $k$. For the coordinate representation $X_a$ in Eq.~(\ref{eq:gmm:means:coo}), apart from $K$, we also need to group the index $C$. For the horizontal representation $X_b$ in Eq.~(\ref{eq:gmm:means:hor}), we need to define the computation of each dimension $d$ in the aggregation expression. As the dimension increases, Eq.~(\ref{eq:gmm:means:coo}) faces the problem of high I/O cost while Eq.~(\ref{eq:gmm:means:hor}) leads to a verbose query. 
Consider the row-major representation which nesting separate dimension attributes into one vector-type attribute. By introducing the $\cdot$ operator for vector computation, Eq.~(\ref{gmm:means}) is expressed in an efficient and clear way (Eq.~(\ref{eq:gmm:means:row})).

\stitle{Relational Algebra to Linear Algebra}.
On the basis of array/vector data type and the derived statistical function and linear algebra operations, the complicated linear algebra computation can be expressed by basic relational algebra operations (selection ($\sigma$),
projection ($\Pi$), union ($\cup$), Cartesian
product ($\times$), and rename ($\rho$)), together with group-by \&
aggregation.
%
%
Let $V$ and $E$ ($E'$) be the relation representation of vector and matrix, such that $V(\TID, v)$ and $E(F, T, e)$. Here $\TID$ is the tuple identifier in $V$. $F$ and $T$, standing for the two indices of a matrix. 
\cite{conf/sigmod/ZhaoY17} introduces two new operations to support the multiplication between a matrix and a vector (Eq.~(\ref{eq:mv})) and between two matrices (Eq.~(\ref{eq:mm})) in their relation representation.  
\begin{eqnarray}
\label{eq:mv}
\displaystyle{E \mathop{\Join}_{T = \TID}^{\oplus(\odot)} V} &=&
_{F}{\mathcal{G}}_{\oplus(\odot)}(E \mathop{\Join}_{T = \TID} V) 
 \\
\label{eq:mm}
\displaystyle{E \mathop{\Join}_{E.T =
    E'.F}^{\oplus(\odot)} E'} &=&
     _{E.F, E'.T}{\mathcal{G}}_{\oplus(\odot)}
     (E \mathop{\Join}_{E.T = E'.F} E')
\end{eqnarray}
The matrix-vector multiplication (Eq.~(\ref{eq:mv})) consists of two steps. The first step is computing $v \odot e$ between a tuple
in $E$ and a tuple in $V$ under the join condition $E.T
= V.\ID$.  The second step is aggregating all the $\odot$ results by
the operation of $\oplus$ for every group-by grouping by the attribute
$E.F$.
%
Similarly, the matrix-matrix multiplication (Eq.~(\ref{eq:mm})) is done in two steps. The first
step computes $\odot$ between a tuple in $E$ and a tuple 
in $E'$ under the join condition $E.T = E'.F$. The second step 
aggregates all the $\odot$ results by the operation of $\oplus$ for
every group by grouping-by the attributes $E.F$ and $E'.T$.
%
The formula of re-estimating the means $\bm{\mu}$ (Eq.~(\ref{gmm:means})) is a matrix-vector multiplication if data is 1-dimensional or a matrix-matrix multiplication otherwise. 
When high dimensional data is in coordinate representation (Table~\ref{tbl:represent}(a)), Eq.~(\ref{eq:gmm:means:coo}) is the application of Eq.~(\ref{eq:mm}). When high dimensional data is nested as the row-major representation (Table~\ref{tbl:represent}(c)), the matrix-matrix multiplication is reduced to matrix-vector multiplication, as shown in Eq.~(\ref{eq:gmm:means:row}).

Re-estimating the covariance/standard deviation $\bm{\sigma}$ (Eq.~(\ref{gmm:covs})) involves the element-wise matrix multiplication if data is 1-dimensional or a tensor-matrix multiplication otherwise. The element-wise matrix multiplication can be expressed by join two matrices on their two indices to compute $E.e \odot E'.e$. An extra aggregation is required to aggregate on each component $k$ as shown in Eq.~(\ref{eq:mm2}).
\begin{eqnarray}
\label{eq:mm2} 
\displaystyle{E \mathop{\Join}_{ \substack{E.F =
    E'.F \\ E.T= E'.T}}^{\oplus(\odot)} E'} &=&
     _{E.F}{\mathcal{G}}_{\oplus(\odot)}
    (E \mathop{\Join}_{ \substack{E.F = E'.F \\ E.T = E'.T}} E')
\end{eqnarray}
Similarly, when $\odot$ and $\oplus$ are vector operation and vector aggregation,  Eq.~(\ref{eq:mm2}) is reduced to high dimensional tensor-matrix multiplication.

\stitle{Value Updating}.
So far, we still need to deal with parameter update when training the model in multiple iterations.  
There is a new relational operation, \UUNION, 
denoted as $\uu$ and first defined in~~\cite{conf/sigmod/ZhaoY17}~(Eq.~(\ref{eq:uu})) to address value update in iterative graph computation.
%
\begin{equation}
\displaystyle{ R \uu_{A} S } =
(R - (R \mathop{\ltimes}_{R.A = S.A} S))\cup S
\label{eq:uu}
\end{equation}
Suppose $t_r$ is a tuple in $R$ and $t_s$ is a tuple in $S$. Different
from the conventional union ($\cup$), the \UUNION updates $t_r$ by
$t_s$ if $t_r$ and $t_s$ are identical by some attributes $A$. 
If $t_s$ does not match any $t_r$, $t_s$ is merged into the resulting relation.
%
Given the relation of parameters as shown in Table~\ref{tbl:rels}(a), the \UUNION updates the set of old parameters by the set of new parameters if they have the identical component identifier $K$.

\stitle{Iterative Evaluation}.
In the following, we elaborate on our approach supporting the iterative model training by \SQL recursive queries.
\begin{figure}[tp!]
{\footnotesize 
\framebox[3.3in]{\begin{minipage}{3.3in}
\mat{0ex}{
\bcc \hspace{1ex} \= {\bf with} \\
\icc \> $TC$ \= $(F, T)$ {\bf as} ( \\
\icc \>\> ({\bf select} \= $F$, $T$ {\bf from}  $E$) \\
\icc \>\>{\bf union all} \hspace{2ex} \\ 
\icc \>\> ({\bf select} \= $TC.F$, $E.T$ {\bf from} $TC$, $E$
       {\bf where} $TC.T = E.F$)) 
}
\vspace{0ex}
\end{minipage}}
\vspace{0ex}
\caption{The recursive \WITH statement}  
\label{sql:tc}
}
\vspace{0ex}
\end{figure}
Over decades, \rdbms have provided the functionality to support
recursive queries, based on \SQL'99
\cite{melton2001sql,Finkelstein96}. 
The recursive queries are expressed using \WITH clause in
\SQL. We introduce the \WITH clause following the discussions given in
\cite{TheCompleteBook}.

\mat{0ex}{
  \hspace{1ex} \= {\bf with} $R$ {\bf as}
  $\langle$ $R$ initialization $\rangle$
  $\langle$ recursive querying involving $R$ $\rangle$ 
}

\noindent
Here, the recursive \WITH clause defines a temporary recursive
relation $R$ in the initialization step, and queries by referring the
recursive relation $R$ iteratively in the recursive step until $R$
cannot be changed. As an example, the edge transitive closure can be
computed using \WITH over the edge relation $E(F, T)$, where $F$ and
$T$ are for ``From'' and ``To''.  As shown in Fig.~\ref{sql:tc}, the
recursive relation is named $TC$. Initially, the recursive relation
$TC$ is defined to project the two attributes, $F$ and $T$, from the
relation $E$ (line~3). Then, the query in every iteration is to union
$TC$ computed and a relation with two attributes $TC.F$ and $E.T$ by
joining the two relations, $TC$ and $E$, over the join condition $TC.T
= E.F$ (line~5).
However, \SQL'99 defines the recursion in a limited form. 
This \WITH clause restricts the recursion to be a stratified program, where
non-monotonic operation, e.g., \UUNION is not allowed.

To support iterative model update, we extend the existing looping structure 
followed the ``algebra + while'' given in \cite{AbiteboulHV95}.

\vspace*{-0.0cm}
\begin{tabbing}
\hspace{2ex} \= initialize $R$ \\
\> {\bf while} \= ($R$ changes) \{ 
$\cdots$; $R \leftarrow \cdots$ \}
\end{tabbing}
\vspace*{-0.0cm}

In brief, in the looping, $R$ may change by the relational algebra in
the body of the looping. The looping will terminate until $R$ becomes
stable.  As discussed in \cite{AbiteboulHV95}, there are two semantics
for ``algebra + while'', namely, noninflationary and inflationary.
The assignment, $R \leftarrow {\mathcal{E}}$, is to assign
relation $R$ by evaluating the relational algebra expression ${\mathcal{E}}$.
For the inflationary semantic, the assignment needs to be cumulative, like the conventional 
union ($\cup$). 
For the noninflationary semantic, the assignment can be destructive, i.g., the new value overwrites the old value, like \UUNION ($\uu$).  
In~\cite{conf/sigmod/ZhaoY17}, it proves that under the framework ``algebra + while'', $\uu$ leads to a fixpoint
in the enhanced recursive \SQL queries by \XY-stratification. 
The vector/matrix data type, just as a nested attribute, does not violate this property in this paper. 
Meanwhile, the \Datalog program of Eq.~(\ref{eq:mm2}) and other relational algebra expressions that consist of natural join, group-by and aggregation can be proved to be \XY--stratification in a similar way as Eq.~(\ref{eq:mv}) and Eq.~(\ref{eq:mm}) in \cite{conf/sigmod/ZhaoY17}. 
%
%

\begin{figure}[t]
{\footnotesize 
\framebox[3.3in]{\begin{minipage}{3.2in}
\mat{0ex}{
  \hspace{1ex} \= {\bf with} \= $R$ {\bf as} \\
  \>\> {\bf select} $\cdots$ {\bf from} $R_{1,j}, \cdots$ {\bf
    computed by} $\cdots$ ($Q_1$) \\
  \>\> {\bf union by update} \\
  \>\> {\bf select} $\cdots$ {\bf from} $R_{2,j}, \cdots$ {\bf
    computed by} $\cdots$ ($Q_2$)  
}
\end{minipage}}
\caption{The general form of the enhanced recursive \WITH}
\label{fig:withE}
}
\end{figure}

To this end, the general syntax of the enhanced recursive \WITH is sketched in Fig.~\ref{fig:withE}. 
In the main body of the enhanced \WITH, apart from \UNION~\ALL, we also allow \UUNION to union the result of initial query $Q_1$ and recursive query $Q_{2}$.
As the discussion in \cite{conf/sigmod/ZhaoY17}, when \UUNION is used, it cannot be used more than once, and cannot be used with other \UNION~\ALL together. 
Here, the \COMPUTEDBY statement, as a new feature of enhanced \WITH, allows users to specify how a relation $R_{i,j}$ is computed by a sequence of queries.  The queries wrapped in \COMPUTEDBY must be non-recursive.
In this paper, we mainly use the \UUNION of the recursive query for parameters update instead of \UNION~\ALL.
In the following sections, we elaborate on the recursive query for EM and model updating in detail. 

\section{Implementation Details}
\label{sec:recforem}
We show the details of supporting the model-based view by the "algebra + while" approach, using \SQL recursive query. 
First, we present the relational algebra expressions needed, followed by the enhanced recursive query and our implementation. Second, the queries for model inference are introduced. 
\subsection{Parameter Estimation}
For simplicity, here we consider the training data point $x_{i}$ is 1-dimensional scalar. It is natural to extend the query to high dimensional input data when matrix/vector data type and functions are supported by the database system.  
We represent the input data by a relation $X(\TID, x)$, where $\TID$ is the tuple identifier for data point $x_{i}$ and $x$ is the numeric value. The model-based view, which is persisted in the relation GMM(K, pie, mean, cov), where $K$ is the identifier of the $k$-th component, and 'pie', 'mean', and 'cov' denote the corresponding parameters, i.e., mixing coefficients,  
means and covariances (standard deviations), respectively. 
The relation representations are shown in Table~\ref{tbl:rels}. 
The following relational algebra expressions describe the E-step~(Eq.~(\ref{eq:algebra:estep})), M-step~(Eq.~(\ref{eq:algebra:mstep1})-(\ref{eq:algebra:mstep4})), and parameter updating (Eq.~(\ref{eq:algebra:update})) in one iteration.

\begin{align}
\displaystyle{ R } & \leftarrow \rho_{(\TID, K, p)} \Pi_{(\TID, K, f)}(GMM \times X) \label{eq:algebra:estep} \\
\displaystyle{ N }  & \leftarrow  \rho_{(K, \text{pie})}( R \mathop{\Join}_{\substack{R.\TID = X.\TID}}^{\SUM(p)} X) \label{eq:algebra:mstep1} \\
\displaystyle{ M } & \leftarrow  \rho_{(K,\text{mean})}( R \mathop{\Join}_{\substack{R.\TID = X.\TID}}^{\SUM(p * x) / \SUM(p)} X) \label{eq:algebra:mstep2} \\
\displaystyle{ T } & \leftarrow   \Pi_{\TID, K, \kw{pow}(x - \text{mean}) }(X \times N) \label{eq:algebra:mstep3} \\ 
\displaystyle{ C } & \leftarrow \rho_{(K, \text{cov})} {_{K}}{\mathcal{G}}_{\SUM(p * t)}(T  \mathop{\Join}_{\substack{R.\TID = T.\TID \\ R.K = T.K}} R) \label{eq:algebra:mstep4} \\
\displaystyle{ GMM } & \leftarrow  \rho_{(K, \text{pie}, \text{mean}, \text{cov})} (N \mathop{\Join}_{N.K = M.K} M \mathop{\Join}_{M.K = C.K} C) \label{eq:algebra:update}
\end{align}

First, in Eq.~(\ref{eq:algebra:estep}), by performing a Cartesian product of GMM and $X$, each data point is associated with the parameters of each component. The responsibilities are evaluated by applying an analytical function $f$ to compute the normalized probability density (Eq.~(\ref{gmm:estep})) for each tuple, which is the E-step. 
The resulted relation $R(\TID, K, p)$ is shown in Fig.~\ref{tbl:rels}(c).
For the M-step, the mixing coefficients 'pie'~(Eq.~(\ref{eq:algebra:mstep1})), the means 'mean'~(Eq.~(\ref{eq:algebra:mstep2})) and the covariances 'cov'~(Eq.~(\ref{eq:algebra:mstep3})-(\ref{eq:algebra:mstep4})) are re-estimated based on their update formulas in Eq.~(\ref{gmm:means})-(\ref{gmm:mixingcoef}), respectively.
In the end, in Eq.~(\ref{eq:algebra:update}), the temporary relations $N$, $M$ and $C$ are joined on attribute $K$ to merge the parameters. The result is assigned to the recursive relation GMM. 
\begin{figure}[t]
\centering
{ \footnotesize 
\framebox[3.4in]{\begin{minipage}{3.4in}
\vspace{-0.5ex}
\mat{0ex}{
\bcc \hspace{2ex} \= {\bf with} \\
\icc \> {GMM}\=($K$, pie, mean, cov) {\bf as} ( \\
\icc \>\> ({\bf select} \= {$K$, pie, mean, cov} {\bf from} {INIT\_PARA}) \\
\icc \>\>{\bf union by update} $K$ \\ 
\icc \>\> ({\bf select} $N.K$, pie/n, mean, \kw{sqrt}(cov/pie) \\
\icc \>\> {\bf from} $N$, $C$ {\bf where} $N.K =C.K$ \\
\icc \>\> {\bf computed by} \\
\icc \>\> $R$($\TID$, $K$, $p$) {\bf as} \= {\bf select} $\TID$, k, {\NORM}(x, mean, cov) * pie / \\ 
\icc \>\>\> ({\SUM}({\NORM}(x, mean, cov) * pie) {\bf over} ({\bf partition by} $\TID$)) \\
\icc \>\>\> {\bf from} GMM, $X$ \\
\icc\>\> $N$($K$, pie, mean) {\bf as} \= {\bf select} $K$, {\SUM}($p$), {\SUM}($p$ * x) / {\SUM}($p$) \\
\icc\>\>\> {\bf from} $R$, $X$ {\bf where} $R.\TID = X.\TID$ \\ 
\icc\>\>\> {\bf group by} $K$ \\
\icc\>\> $C$($K$, cov) {\bf as} \= {\bf select} $R.K$, {\SUM}($p$ * $T$.val) {\bf from} \\
\icc\>\>\> ({\bf select} $\TID$, $K$, pow(x-mean) {\bf as} val {\bf from} $X$, $N$) {\bf as} $T$, $R$ \\
\icc\>\>\> {\bf where} $T.\TID = R.\TID$ {\bf and} $T.K = R.K$ \\
\icc\>\>\> {\bf group by} $R.K$) \\
\icc \>\> {\bf maxrecursion} 10) \\
\icc \> {\bf select} * {\bf from} {GMM}
}
\end{minipage}}
\caption{ The enhanced recursive \SQL for Gaussian Mixtures}  
\label{with+:gmm}
}
\end{figure}

Fig.~\ref{with+:gmm} shows the enhanced \WITH query to support Gaussian Mixture Model by EM algorithm. The recursive relation GMM specifies the parameters of $k$ Gaussian distributions. 
In line 3, the initial query loads the initial parameters from relation INI\_PARA. 
The new parameters are selected by the recursive query (line 5-6) evaluated by the \COMPUTEDBY statement and update the recursive relation by \UUNION w.r.t. the component index $K$. 
It wraps the queries to compute E-step and M-step of one iteration EM. 

We elaborate on the queries in the \COMPUTEDBY statement (line 8-17). 
Specifically, the query in line 8-10 performs the E-step, as the relational algebra in Eq.~(\ref{eq:algebra:estep}). 
Here, \NORM is the Gaussian (Normal) probability density function of data point $x$ given the mean and covariance as input.
We can use the window function, introduced in \SQL'03 to compute the responsibility by Bayes rule in Eq.~(\ref{gmm:estep}).
In line 9, $\SUM ()~\kw{over}~(\PARTITIONBY()) $ is the window function performing calculation across a set of rows that are related to the current row. 
As it does not group rows, where each row retains its separate identity, many \rdbms allow to use it in the recursive query, e.g., \PostgreSQL and \Oracle.
The window function partitions rows of the Cartesian product results in partitions of the same $\TID$ and computes the denominator of Eq.~(\ref{gmm:estep}).
In line 11-13, the query computes the means (Eq.~(\ref{gmm:means})) and the mixing coefficients together by a matrix-matrix multiplication due to their common join of $R$ and $X$. 
Then, line 14-17 computes the covariances of Eq.~(\ref{gmm:covs}). 
First, we compute the square of $x_i - \bm{\mu}_{k}$ for each $x_{i}$ and $k$, which requires a Cartesian product of $N$ and $R$ (Eq.~(\ref{eq:algebra:mstep3})). Second, the value is weighted by the responsibility and aggregated as specified in Eq.~(\ref{eq:algebra:mstep4}).
The new parameters in the temporary relation $N$ and $C$ will be merged by joining on the component index $K$ in line 6.

An acute reader may find that in Fig.~\ref{with+:gmm}, the recursive query does not compute and check the convergence of the log-likelihood explicitly.  
That is because the existing recursive query does not support the functionality of checking value convergence as well as early stopping. However, the depth of recursion can be controlled by \MAXREC clause, which is adapted from  \SQLServer~\cite{sqlserverdocs}. 
The \MAXREC clause can effectively prevent infinite recursion because of infinite fix point, e.g.,  '
  {\bf with} $R(n)$ {\bf as} 
  (({\bf select} {\bf values}(0)) {\bf union all} ({\bf select} $n
  + 1$ {\bf from} $R$)) '
, a standard \SQL'99 recursion.
Users can check the convergence after training for a fixed number of recursion and resume the training  from current parameters if necessary.

\stitle{The implementation}: We sketch how to support recursive queries
using the enhanced \WITH in \rdbms. 
First, for each subquery $Q_i$ used in $Q$ including those defined by
the \COMPUTEDBY statement, we construct a local dependency graph
$G_i$. The graph $G_i$ constructed must be cycle free.
We ensure that it is \XY-stratified.
Second, we create a \PSM (Persistent Stored Model) in the recent \SQL
standard.
With \PSM, we create a unique procedure $F_Q$ for the recursive query
$Q$ to be processed, as illustrated below.

\mat{0ex}{
  \hspace{1ex} \= {\bf cr}\={\bf eate procedure} $F_Q$ ( \\
  \>\>  {\bf declare} $C_1, \cdots, C_i, \cdots$; \\
  \>\>  {\bf create table} $R_{i,j}$ for all tables defined by \kw{as}
        in a subquery $Q_i$; \\
        \>\>create \SQL statement to compute the initial $R$ by union of \\
        \>\>all initial subqueries; \\  
\>\> {\bf lo}\={\bf op} \\
\>\>\>  {\bf insert into} $R_{i,j}$ {\bf select} $\cdots$ for every
$R_{i,j}$ used in $Q_i$; \\
\>\>\>  compute condition $C_i$ for each recursive subquery $Q_i$; \\
\>\>\> {\bf if} \=  all $C_i$ for the recursive subqueries are false {\bf
  then exit} \\
\>\>\> compute the recursive relation $R$ for the current iteration;
\\
\>\>\> union the current $R$ with the previous $R$ computed; \\
\>\> {\bf end loop})    
}

\noindent
In the procedure, $F_Q$, first we declare variables $C_1, \cdots, C_i,
\cdots$ for every subquery $Q_i$, which are used to check the condition
to exit from the looping. Second, we create the temporary tables for
the relations defined by \kw{as} in the \COMPUTEDBY statements.
Third, we include \SQL statements to compute the initial recursive
relation $R$. Fourth, we create a looping. In the looping, we generate an \INSERT for 
$R_{i,j}$, and check whether $Q_i$ is empty. 
If so, 
$C_i$ is set to $0$, indicating $Q_i$ generates $0$ tuple, 
the loop will be terminated. Otherwise, the
recursive relation computed in this iteration will union with the
one computed in the previous iteration by either \UNION~\ALL or
\UUNION. An extra counter will be maintained in the loop if \MAXREC is used.  
With the procedure defined, we can run the statements in the
procedure $F_Q$ by issuing ``\kw{call} $F_Q$''.

\subsection{Model Inference}
Once the model is trained by the recursive query in Fig.~\ref{with+:gmm}, it can be materialized in a view for online inference. In the phase of inference, users can query the view by \SQL to perform  clustering, classification and density estimation. 
Given a batch of data in relation $X$ and a view GMM computed by Fig.~\ref{with+:gmm}. The query below computes the posterior probability that the component $K$ generated the data with index $\TID$. The query is similar to computing the E-step (Eq.~(\ref{gmm:estep})) in line 5-7 of Fig.~\ref{with+:gmm}. 
\mat{0ex}{
 \hspace{1ex} \= {\bf create table} $R$ {\bf as} \= {\bf select} $\TID$, $K$, \\
 \> {\NORM}(x, mean, cov) * pie / ({\SUM}({\NORM}(x, mean, cov) * pie) \\
 \> {\bf over} ({\bf partition by} $\TID$)) {\bf from} GMM, $X$
}
Based on relation $R(\TID, K, p)$ above, we can further assign the data into $K$ clusters, where $x_{i}$ is assigned to cluster $k$ if the posterior probability $p(z_{ik})$ is the maximum among the $\{p(z_{i1}, \cdots p(z_{iK})\}$. The query below creates a relation CLU($\TID$, $K$) to persist the clustering result where $\TID$ and $K$ are the attributes of data point and its assigned cluster, respectively. 
It first finds the maximum $p(z_{ik})$ for each data point by a subquery on relation $R$. The result is renamed as $T$ and is joined with $R$ on the condition of $R.\TID = T.\TID$ and $R.p = T.p$ to find the corresponding $k$.

\mat{0ex}{
 \hspace{1ex} \= {\bf create table} CLU {\bf as} \= {\bf select} $\TID$, $K$ {\bf from} $R$, \\
 \> ({\bf select} $\TID$, \kw{max}($p$) {\bf as} $p$ {\bf from} $R$ {\bf group by} $\TID$) {\bf as} $T$, \\
 \> {\bf where} $R.\TID = T.\TID$ {\bf and} $R.p = T.p$ 
}

It is worth nothing that both of the queries above only access the data exactly once. Thereby, it is possible to perform the inference on-the-fly and only for interested data. 
Besides from density estimation and clustering, result evaluation, e.g., computing the purity, normalized mutual information (NMI) and Rand Index can be conducted in database by \SQL queries.

\section{Model Maintenance}
\label{sec:maintain}
\begin{figure}[t]
\centering
 \includegraphics[width=0.8\columnwidth]{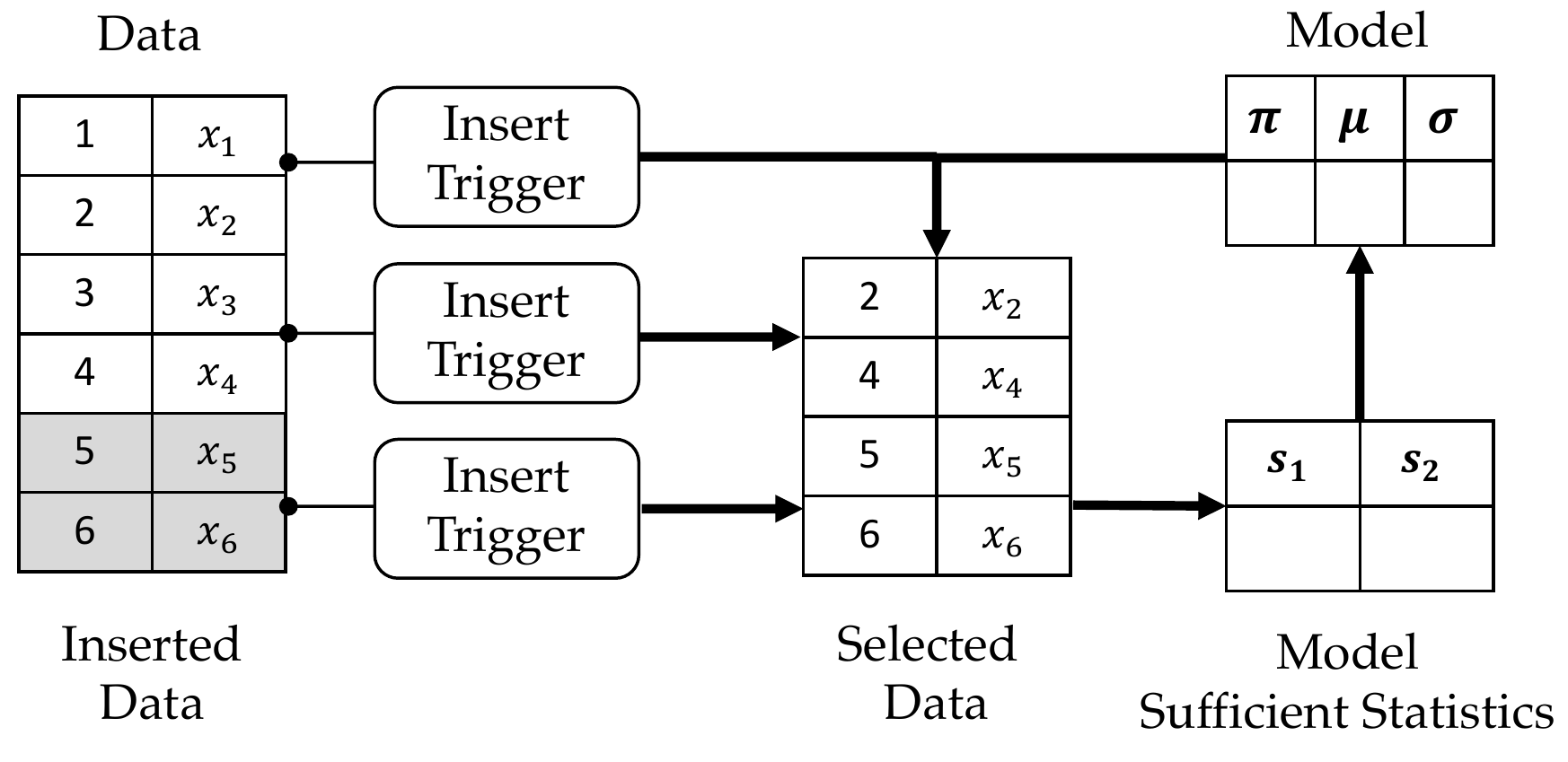} 
 \caption{Overview of Model Maintenance}
  \label{fig:gmm:modelupdate}
\end{figure}

\begin{figure}[t]
\centering
{ \footnotesize 
\framebox[2.8in]{\begin{minipage}{3.6in}
\vspace{-0.5ex}
\mat{0ex}{
\bcc \hspace{2ex} \= {\bf create trigger} T1 {\bf before insert on} $X$ \\
\icc \> \= {\bf for each statement} \\
\icc \> \= {\bf execute procedure} \kw{DATA\_SELECTION} \\ 
 \\ 
\icc \> \= {\bf create trigger} T2 {\bf before insert on} $X$ \\
\icc \> \= {\bf for each row} \\
\icc \> \= {\bf execute procedure} \kw{DATA\_INSERTION} \\ 
 \\ 
\icc \> \= {\bf create trigger} T3 {\bf after insert on} $X$ \\
\icc \> \= {\bf for each statement} \\
\icc \> \= {\bf execute procedure} \kw{MODEL\_UPDATE} 
}
\end{minipage}}
\caption{ The triggers for incremental update}  
\label{fig:triggers}
}
\end{figure}

In this section, we investigate the automatic model/view updating.
When the underlying data $X$ changes, a straightforward way is to re-estimate the model over the updated data. 
However, when only a small portion of the training data are updated, the changes of the corresponding model are slight, it is inefficient to re-estimate the model on-the-fly. 
Hence, a natural idea is arose that whether we can update existing model by exploring the 'incremental variant' of the EM algorithm.  And this variant can be maintained by the newly arriving data and a small portion of data extracted from original dataset.
As the statistical model trained by the \SQL queries can be represented by its sufficient statistics, the model is updated by maintaining the model and sufficient statistics. 

The sufficient statistic is a function of data $X$ that contains all of the information relevant to estimate the model parameters. 
The sufficient statistics of these data are computed to update the statistics of model. As the model is updated, the statistics of data is also updated followed by the changing of the posterior probability $p(z_{ik})$.  This process repeats until the statistics converge. 
It is worth mentioning that this maintenance mechanism can support all the models with have sufficient statistics. 
We elaborate the sufficient statistics updating rules in the following.

Suppose the training dataset of model $\bm{\theta}$ is $\{ x_1, x_2, \cdots, x_n\}$. Let $\bm{s}$ be the sufficient statistics of $\bm{\theta}$, based on the Factorization Theorem~\cite{DudaHartStork01}, we can obtain  
\begin{equation}
\bm{s} = \sum_{i = 1}^{n} \sum_{\bm{z}} p(\bm{z}| x_i, \bm{\theta}) \phi(x_i, \bm{z})
\end{equation}
where $\bm{z}$ is the unobserved variable,  $\phi$ denotes the mapping function from an instance $(x_i, \bm{z})$ to the sufficient statistics contributed by $x_i$. The inserted data is $\{ x_{n+1}, x_{n+2}, \cdots, x_m\}$. Let the new model for overall data $\{ x_1, \cdots, x_n, x_{n+1}, \cdots, x_m\}$ be $\bm{\widetilde{\theta}}$ and the corresponding sufficient statistics be  $\bm{\widetilde{s}}$. The difference of $\bm{\widetilde{s}} - \bm{s}$, denoted as $\Delta\bm{s}$ is 
\begin{align}
\Delta\bm{s} 
& = \sum_{i = 1}^{n+m} \sum_{\bm{z}} p(\bm{z}| x_i, \bm{\widetilde{\theta}}) \phi(x_i, \bm{z}) -  \sum_{i = 1}^{n} \sum_{\bm{z}} p(\bm{z}| x_i, \bm{\theta}) \phi(x_i, \bm{z}) \nonumber \\ 
& = \sum_{i = 1}^{n+m} \sum_{\bm{z}} [p(\bm{z}| x_i, \bm{\widetilde{\theta}}) - p(\bm{z}| x_i, \bm{\theta})] \phi(x_i, \bm{z})  \label{eq:ss:part1} \\
& + \sum_{i = n+1}^{m} \sum_{\bm{z}} p(\bm{z}| x_i, \bm{\theta}) \phi(x_i, \bm{z}) \label{eq:ss:part2}
\end{align} 
%
%
According to above equations, we observe that the delta part of the sufficient statistics $\Delta\bm{s}$ consists of two parts: (1) changes of the sufficient statistics for the overall data points $\{x_{1}, x_{2} \cdots x_{m}\}$ in Eq.~(\ref{eq:ss:part1}), and (2) the additional sufficient statistics for the newly inserted data points $\{x_{n+1}, \cdots x_m\}$ in Eq.~(\ref{eq:ss:part2}).
Consider to retrain a new model $\bm{\widetilde{\theta}}$ over $\{ x_1, x_2, \cdots, x_m\}$  in $T$ iterations by taking $\bm{\theta}$ as the initial parameter, i.e., $\bm{\theta}^{(0)} = \bm{\theta}$ and $\bm{\theta}^{(T)} = \bm{\widetilde{\theta}}$. We have 
\begin{align}
\label{eq:modelupdate:suffstat} 
\Delta\bm{s} 
& = \sum_{i = 1}^{n+m} \sum_{\bm{z}} [p(\bm{z}| x_i, \bm{{\theta}}^{(T)}) - p(\bm{z}| x_i, \bm{\theta}^{(0)})] \phi(x_i, \bm{z}) \\
& + \sum_{i = n+1}^{m} \sum_{\bm{z}} p(\bm{z}| x_i, \bm{\theta}^{(0)}) \phi(x_i, \bm{z})  \\
& = \sum_{t = 1}^{T} \sum_{i = 1}^{n+m} \sum_{\bm{z}}  [p(\bm{z}| x_i, \bm{{\theta}}^{(t)}) - p(\bm{z}| x_i, \bm{\theta}^{(t-1)})] \phi(x_i, \bm{z})  \label{eq:compute:part1} \\
& + \sum_{i = n+1}^{m} \sum_{\bm{z}} p(\bm{z}| x_i, \bm{\theta}^{(0)}) \phi(x_i, \bm{z}) \label{eq:compute:part2}
\end{align} 
Above equations indicts how to compute $\Delta\bm{s}$. 
For the inserted data $\{x_{n+1}, \cdots x_m\}$, the delta can be directly computed by evaluating the original model $\bm{\theta}^{(0)}$ as Eq.~(\ref{eq:compute:part2}). 
while for original data, the delta can be computed by updating the model $\bm{\theta}^{(t)}$ iteratively using all the data $\{x_{1}, x_{2} \cdots x_{m}\}$ as Eq.~(\ref{eq:compute:part1}). Since most of the computational cost is concentrated on the iteration of Eq.~(\ref{eq:compute:part1}), we use two tricks to approximate the computation.   
First, we use the stochastic approximation algorithm, where the parameters are updated after the sufficient statistics of each new data point $x_{i}$ is computed, instead of the full batch dataset. This approach is widely used in many online and incremental EM algorithm variations~\cite{titterington1984recursive, DBLP:books/sp/12/NealH98, DBLP:conf/naacl/LiangK09}. 
The second is discarding the data points which are not likely to change their cluster in the future, as the scaling clustering algorithms adopt for speedup~\cite{DBLP:conf/kdd/BradleyFR98}. 
In other words, due to the slight changes of data and models, only a small portion of the original data points with unstable membership need to be retrained.
We discuss our strategy of selecting partial original data in $\{ x_1, x_2, \cdots, x_n\}$ for model update. It is a tradeoff between the accuracy of the model and the updating cost. The more data we use to update the model, the more accuracy the model we can achieve while the larger computation and I/O cost to spend. The data points which are unlikely to change cluster membership can be discarded in the future model update. 
There are two strategies for the discard, a distance-based and a density-based strategy.
For the distance-based strategy, we use Mahalanobis distance~\cite{DBLP:books/lib/DudaH73} to measure the distance between a data point and a distribution. For each data $x_{i}$, we compute the Mahalanobis distance, $D_{k}(x_{i})$, to the $k$-th component with mean $\bm{\mu}_{k}$ and covariance $\bm{\sigma}_{k}$.

\begin{equation}
D_{k}(x_{i}) = \sqrt{(x_{i} - \bm{\mu}_k)^{T}\bm{\sigma}_{k}^{-1}(x_{i} - \bm{\mu}_{k})}
\end{equation}

We can filter the data within a given thresholding radius with any component.
Another measurement is the entropy of the posterior probability for data $x_i$ as in Eq.~(\ref{eq:entropy}), where $p(z_{ik})$ is evaluated by parameter $\bm{\theta}^{(0)}$. The larger the entropy, the lower the possibility of assigning $x_{i}$ to any one of the component. 
\begin{equation}
\label{eq:entropy}
E(x_{i}) = - \sum_{k = 1}^{K} p(z_{ik}) ln~p(z_{ik})
\end{equation}
It is worth mentioning that the data selection in trigger \kw{T1} can be performed offline, i.e., persisting a subset of training data with fixed budget size for model updating in the future. In addition, the sufficient statistics for original model $\bm{\theta}^{0}$ (line 1 of Algorithm~\ref{algo:modelupdate}) can be precomputed. Those will improve the efficiency of online model maintenance significantly.  The alternative fine-grained data selection strategies are beyond the scope of this paper.

 Similarly, considering deleting $m$ data points $\{x_{n-m+1}, \cdots x_n\}$ from $\{x_{1}, x_{2} \cdots x_{n}\}$, the difference of the sufficient statistics, $\Delta \bm{s}$ is 
\begin{align} 
\Delta \bm{s} & = \sum_{t = 1}^{T} \sum_{i = 1}^{n-m} \sum_{\bm{z}}  [p(\bm{z}| x_i, \bm{\theta}^{(t)}) - p(\bm{z}| x_i, \bm{\theta}^{(t-1)})] \phi(x_i, \bm{z}) \\ \nonumber
& -\sum_{i = n-m+1}^{n} \sum_{\bm{z}} p(\bm{z}| x_i, \bm{\theta}^{(0)}) \phi(x_i, \bm{z})
\end{align} 

In \rdbms, the automatic model updating mechanism is enabled by triggers build on the relation of the input data $X$.
Fig.~\ref{fig:gmm:modelupdate} illustrates the overview of our model/view updating mechanism.  
There are three triggers built on the relation of training data $X$, whose definitions are shown in Fig.~\ref{fig:triggers}.
%
Before executing the insertion operation, two triggers \kw{T1}~(line 1-3 in Fig.~\ref{fig:triggers}) and \kw{T2}~(line 4-6 in Fig.~\ref{fig:triggers})  prepare the data for model updating in a temporary relation $X'$. Here, \kw{T1} performs on each row to select a subset from original data in $\{ x_1, x_2, \cdots, x_n\}$ based on a selection criterion. Additionally,  \kw{T2} inserts all the newly arrived data  $\{ x_{1+n}, x_2, \cdots, x_m\}$ to relation  $X'$. 
After the data preparation finished, another trigger \kw{T3}~(line 7-9 in Fig.~\ref{fig:triggers}) will call a \PSM to compute the $\Delta \bm{s}$ by $X'$. 
In the \PSM, first, the delta of the newly inserted data (Eq.~(\ref{eq:compute:part1})) is computed as used to reinitialize the parameters of the model. Then, $T$ iterations of scanning relation $X'$ is performed. Where in each iteration. $X'$ is randomly shuffled and each data point is used to update the sufficient statistics it contributes as well as the model instantly. 
The actions of these triggers are transparent to the database users. 
Finally, we use Gaussian Mixture model an example to illustrate this procedure.  
\begin{example}
For Gaussian Mixture model of $K$ components, the minimal sufficient statistics $\bm{s} = ({s_{11}, s_{21}, \cdots, s_{1K}, s_{2K}})$, where $s_{1k} \in \mathbb{R}$ and $s_{2k} \in \mathbb{R}^{d \times d}$ for each $k \in \{ 1, \cdots, K \}$ are as below.
\begin{equation}
\label{eq:gmm:suffstat}
s_{1k} = \sum_{i = 1}^{n} p(z_{ik}) x_i, ~~~~
s_{2k} = \sum_{i = 1}^{n} p(z_{ik}) x_{i} x_{i}^T
\end{equation}
And the parameter can be computed by the sufficient statistics as shown in Eq.~(\ref{eq:gmm:compute:suffstat}). 
\begin{align}
\label{eq:gmm:compute:suffstat}
\bm{\mu}_k & = \frac{1}{n} s_{1k}, ~~ \bm{\pi}_k = \frac{1}{n} \sum_{i = 1}^{n}p(z_{ik}) \\ 
\bm{\sigma}_k & =  \frac{1}{n} s_{2k} - \bm{\mu}_k \bm{\mu}_k^T \nonumber
\end{align}
When an insert command of relation $X$ is issued, at first, trigger \kw{T1} and \kw{T2} prepare the temporary relation $X'$. \kw{T3} is triggered followed by the insertion command. 
The procedure \kw{T3} executed is illustrated in Algorithm~\ref{algo:modelupdate}. 
In line 2, tigger \kw{T3} first adds the sufficient statistics of the inserted data into $\bm{s}$. Then it further updates $\bm{s}$ by performing $T$ iterations over $X'$, in each iteration, each data point $x_i$ is used to update $\bm{s}$ as well as the model  instantly. Here, $p(z_{ik})^{(t)}$ is the responsibility of $x_i$ (Eq.~(\ref{gmm:estep})) evaluated in $t$-th iteration.
\comment{
The called \PSM first adds the sufficient statistics of the inserted data into $\bm{s}$ by Eq.~(\ref{eq:gmm:suffstat}). Then it further updates $\bm{s}$ by performing $T$ iterations over $X'$, where specifically, in each iteration, each $x_i$ is used to update $\bm{s}$ by Eq.~(\ref{eq:gmm:suffstatupdate}) instantly. 
\begin{align}
\label{eq:gmm:suffstatupdate}
s_{1k} & \leftarrow  s_{1k} +  p(z_{ik})^{(t)} x_i - p(z_{ik})^{(t - 1)} x_i \\ 
s_{2k} & \leftarrow  s_{2k} +  p(z_{ik})^{(t)} x_{i} x_{i}^T - p(z_{ik})^{(t - 1)} x_{i} x_{i}^T \nonumber
\end{align}
Meanwhile, the parameters of the model are re-estimated by Eq.~(\ref{eq:gmm:compute:suffstat}) and updated. 
}
\qed 
\end{example}

\begin{algorithm}[t]
\caption{{\kw{MODEL\_UPDATE}}}
\label{algo:modelupdate}
\begin{algorithmic}[1]
\STATE  Initialize the original sufficient statistics $\bm{s}$ by Eq.~(\ref{eq:gmm:suffstat});
\STATE  $s_{1k} \leftarrow s_{1k} + \sum_{i = n+1}^{m} p(z_{ik}) x_i$; $s_{2k} \leftarrow s_{2k} + \sum_{i = n + 1}^{m} p(z_{ik}) x_{i} x_{i}^T $;
\STATE  Update model parameters $\bm{\mu}$, $\bm{\pi}$ and $\bm{\sigma}$ by Eq.~(\ref{eq:gmm:compute:suffstat});
\FOR{$t \gets 1$ to $T$} 
	\FOR {$x_i \in X'$ in random order}
		\STATE {$s_{1k} \leftarrow  s_{1k} +  p(z_{ik})^{(t)} x_i - p(z_{ik})^{(t - 1)} x_i$; 
				$s_{2k} \leftarrow  s_{2k} +  p(z_{ik})^{(t)} x_{i} x_{i}^T - p(z_{ik})^{(t - 1)} x_{i} x_{i}^T$;}
		\STATE  Update model parameters $\bm{\mu}$, $\bm{\pi}$ and $\bm{\sigma}$ by Eq.~(\ref{eq:gmm:compute:suffstat});
	\ENDFOR
\ENDFOR 
\RETURN $\bm{\mu}$, $\bm{\sigma}$, $\bm{\pi}$;
\end{algorithmic}
\end{algorithm}

\section{Experimental Studies}
\label{sec:exp}
In this section, we present our experimental studies of supporting model-based view training, inference, and maintenance in \rdbm. We conduct extensive experiments to investigate the following facets:
\begin{itemize}
\item compare the performance of our enhanced \WITH and looping control by a host language.
\item test the scalability of the recursive queries for different models on synthetic data.
\item conduct a case study on market segmentation on retail data.
\item validate the efficiency of our model maintenance mechanism. 
\end{itemize}

\stitle{Experimental Setup}: We report our performance studies on a PC with 
Intel(R) Xeon(R) CPU E5-2697 v3 (2.60GHz) with 96GB RAM running Linux CentOS 7.5 64 bit. 
We tested the enhanced recursive query on \PostgreSQL 10.10~\cite{postgresdocs}. 
The statistical function and matrix/vector computation function are supported by Apache MADlib 1.16~\cite{DBLP:journals/pvldb/HellersteinRSWFGNWFLK12}.
All the queries we tested are evaluated in a single thread \PostgreSQL instance.
\subsection{\WITHplus vs. \Psy}
\begin{figure}[t]
\centering
\subfigure[varying d]{
\vspace*{-0.4cm}
\label{fig:exp:python:dim}
 \includegraphics[width=0.4\columnwidth]{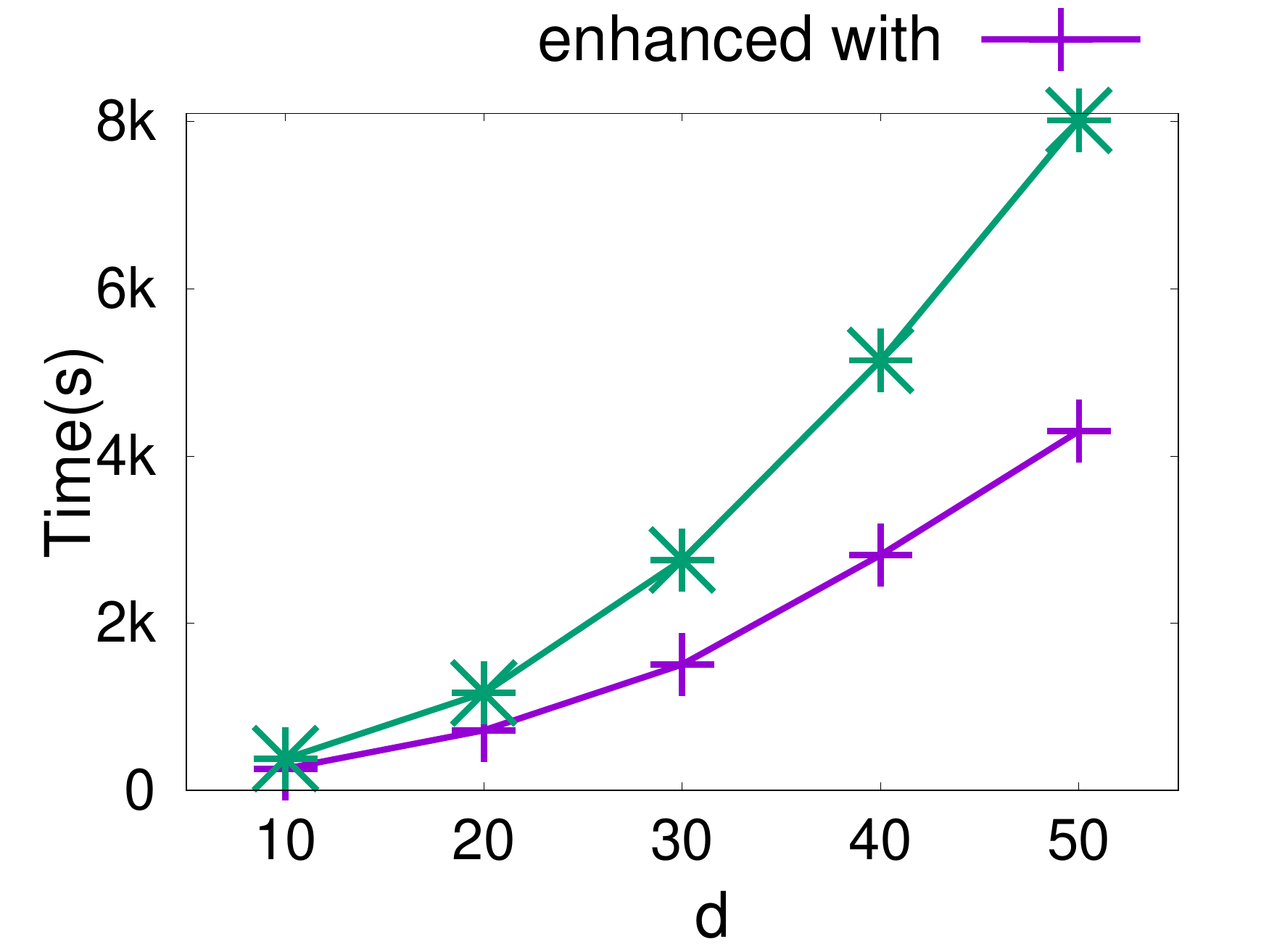}
}
\subfigure[varying n]{
\label{fig:exp:python:num}
\vspace*{-0.4cm}
 \includegraphics[width=0.4\columnwidth]{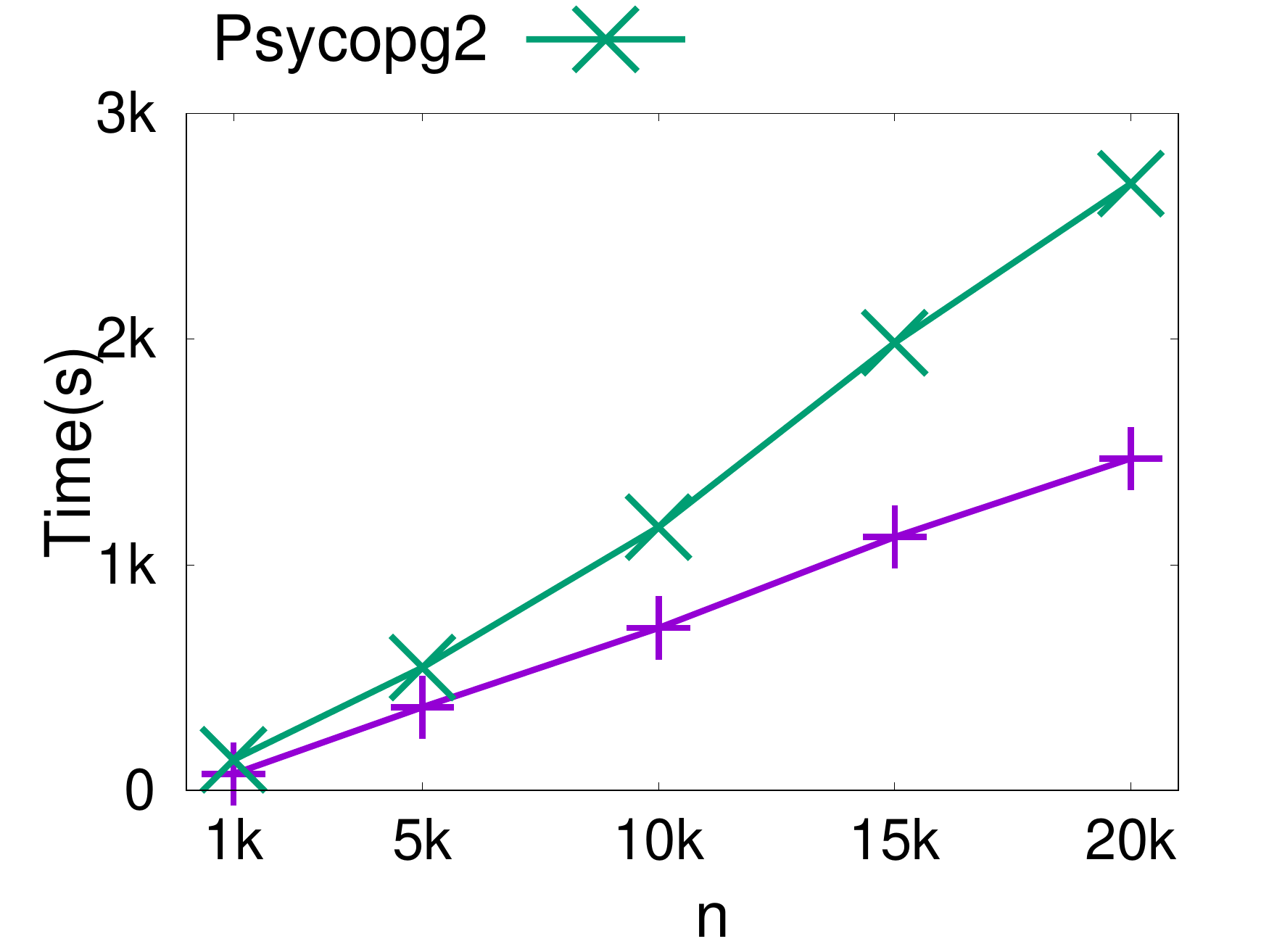}
}
\subfigure[varying k]{
\vspace*{-0.4cm}
\label{fig:exp:python:k}
 \includegraphics[width=0.4\columnwidth]{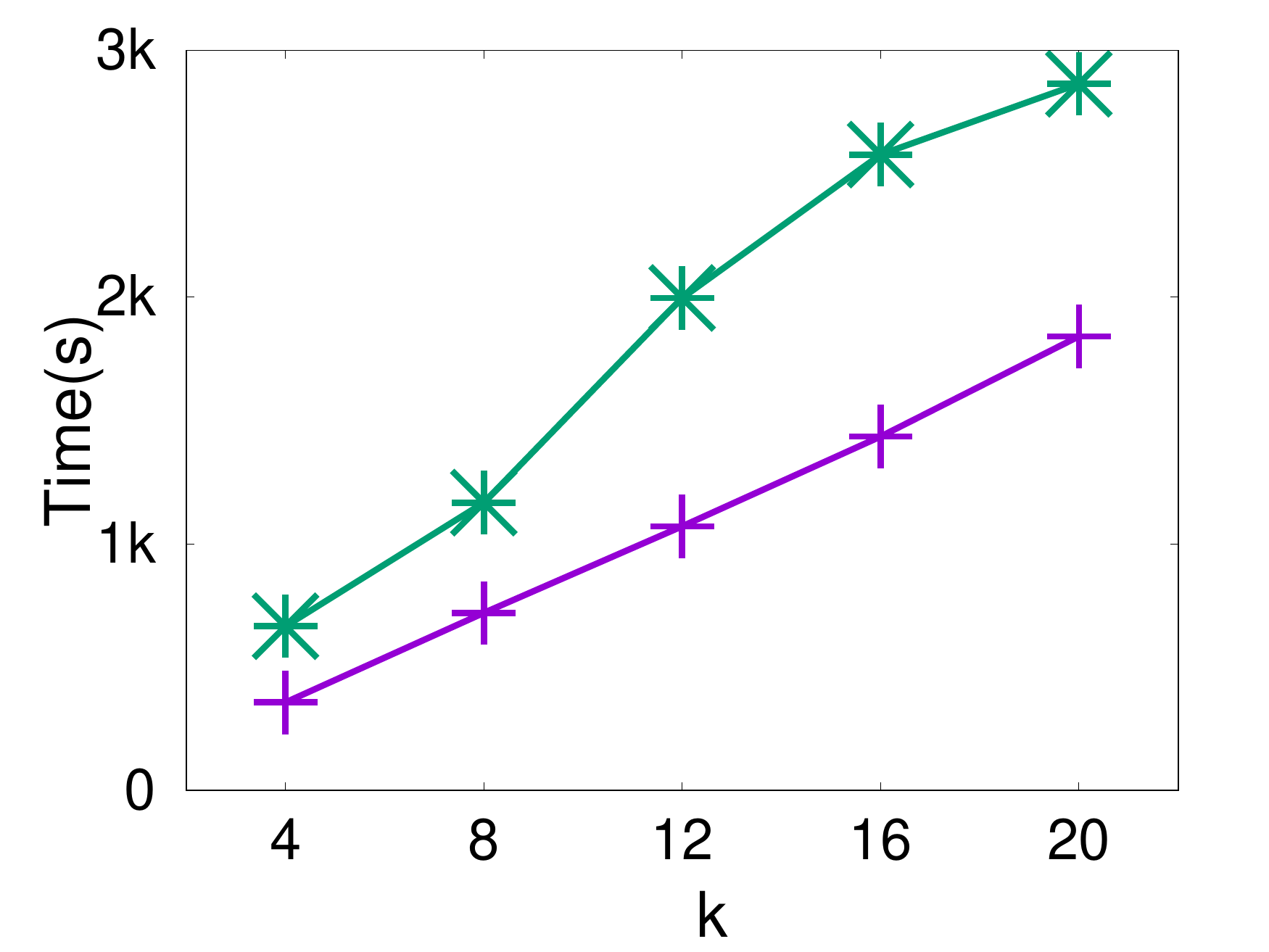}
}
\subfigure[varying number of iterations]{
\vspace*{-0.4cm}
\label{fig:exp:python:iter}
 \includegraphics[width=0.4\columnwidth]{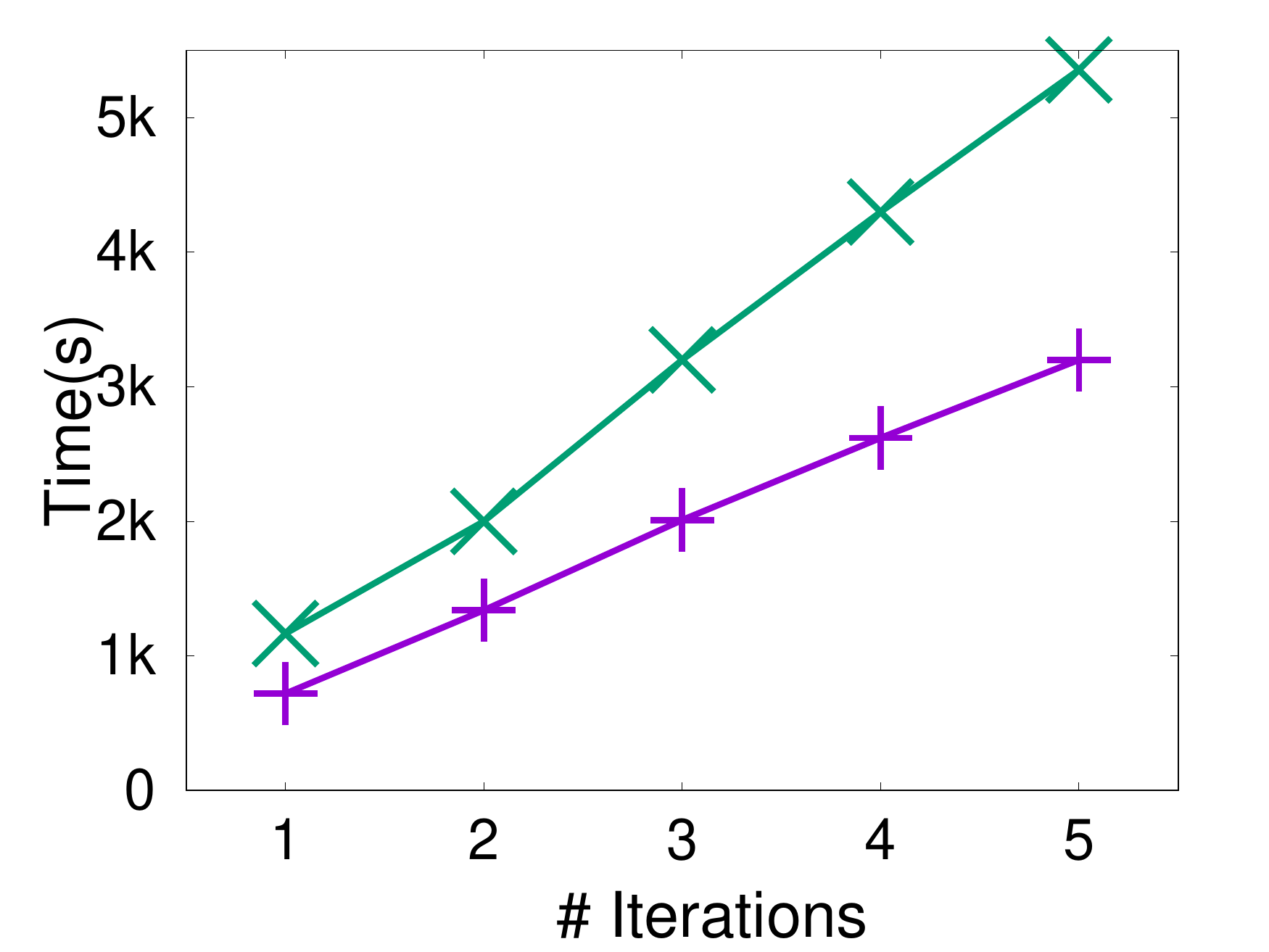}
}
\vspace*{-0.0cm}
\caption{\WITHplus vs. \Psy}
\label{fig:exp:python}
\vspace*{-0.0cm}
\end{figure}

\begin{figure*}[t]
\centering
\subfigure[k = 8, n = 10000, varying d]{
\vspace*{-0.4cm}
\label{fig:exp:scale:d}
 \includegraphics[width=0.55\columnwidth]{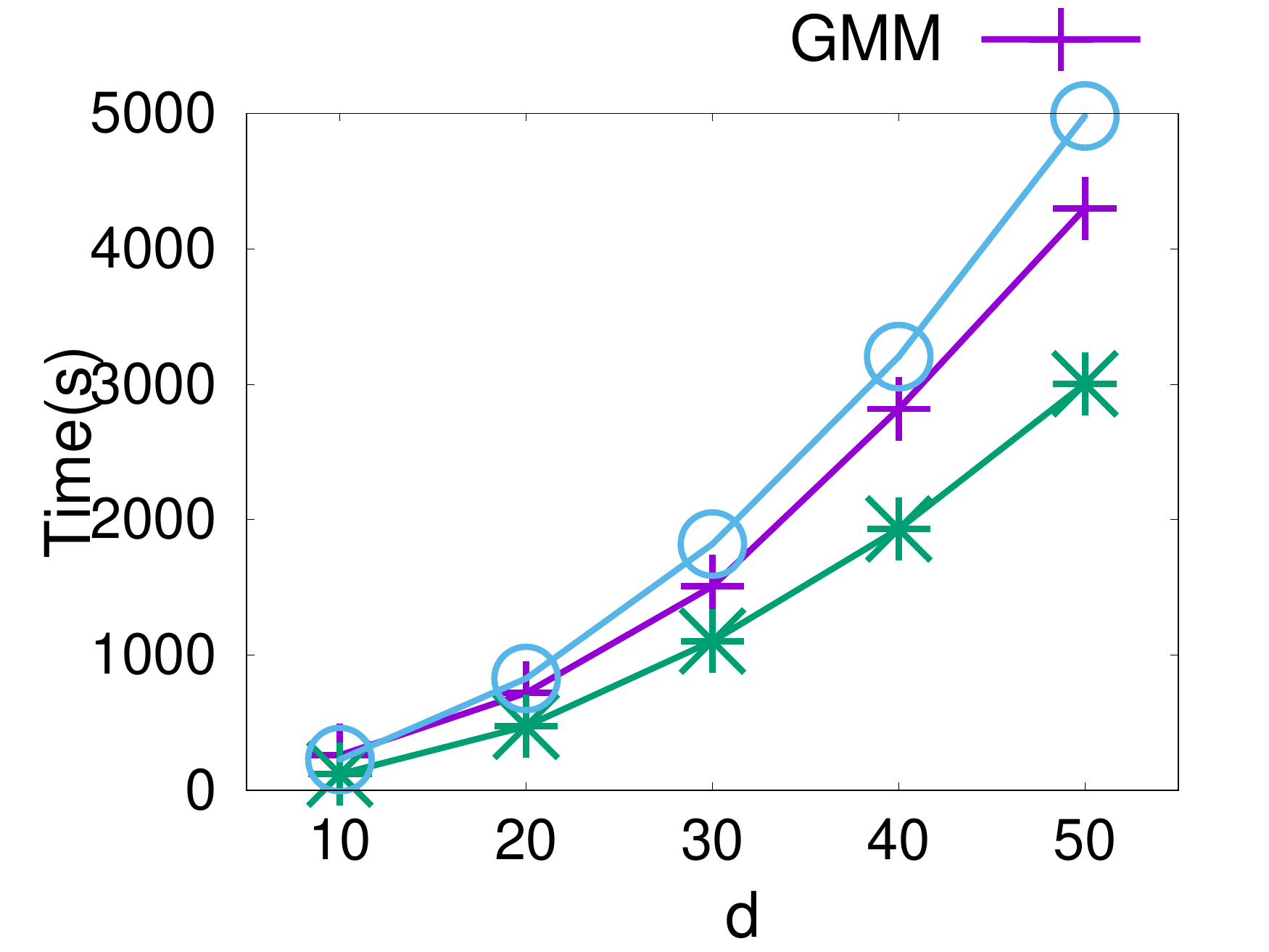}
}
\subfigure[k = 8, d = 20, varying n]{
\label{fig:exp:scale:n}
\vspace*{-0.4cm}
 \includegraphics[width=0.55\columnwidth]{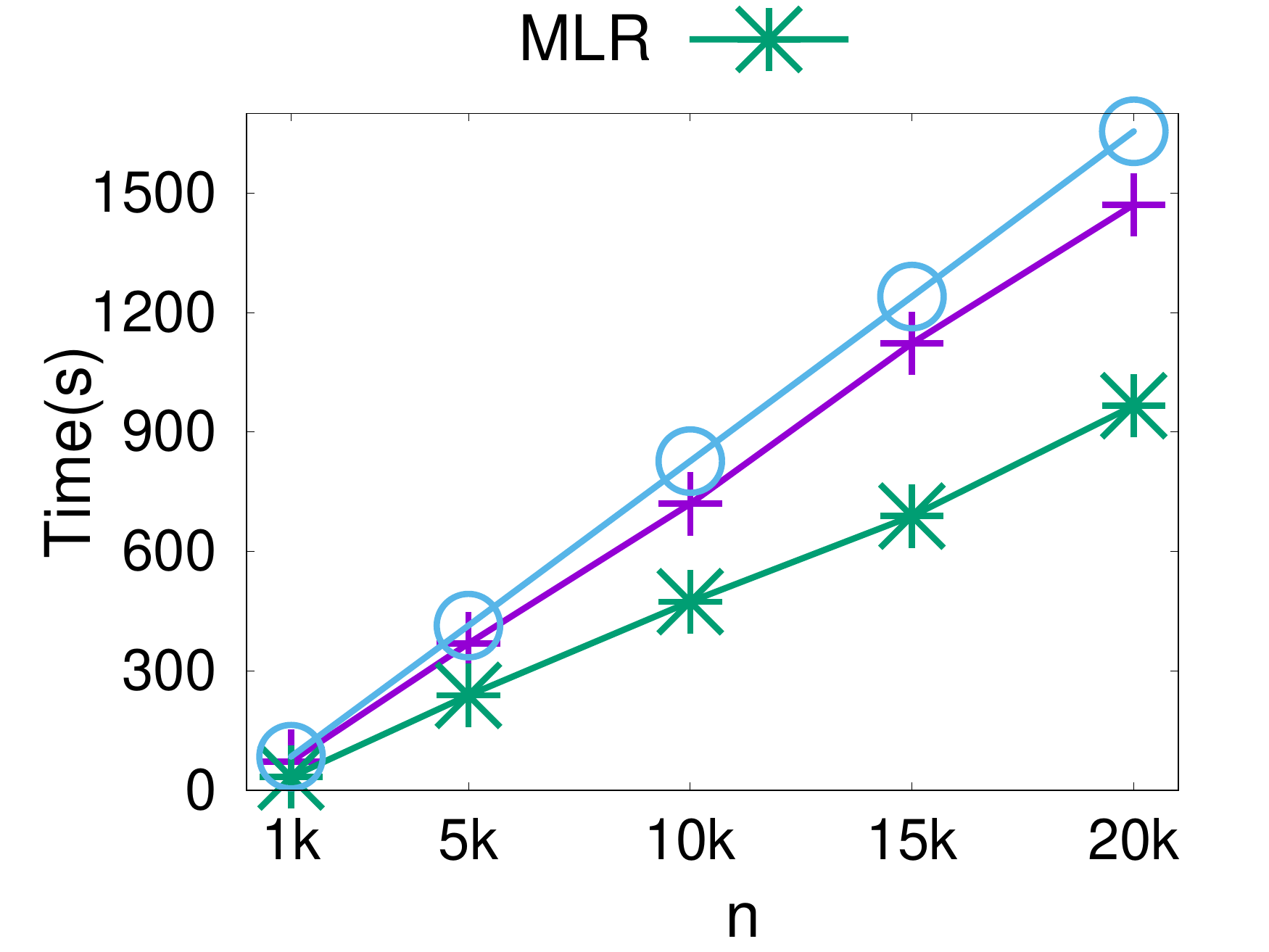}
}
\subfigure[d = 20, n = 10000, varying k]{
\vspace*{-0.4cm}
\label{fig:exp:scale:k}
 \includegraphics[width=0.55\columnwidth]{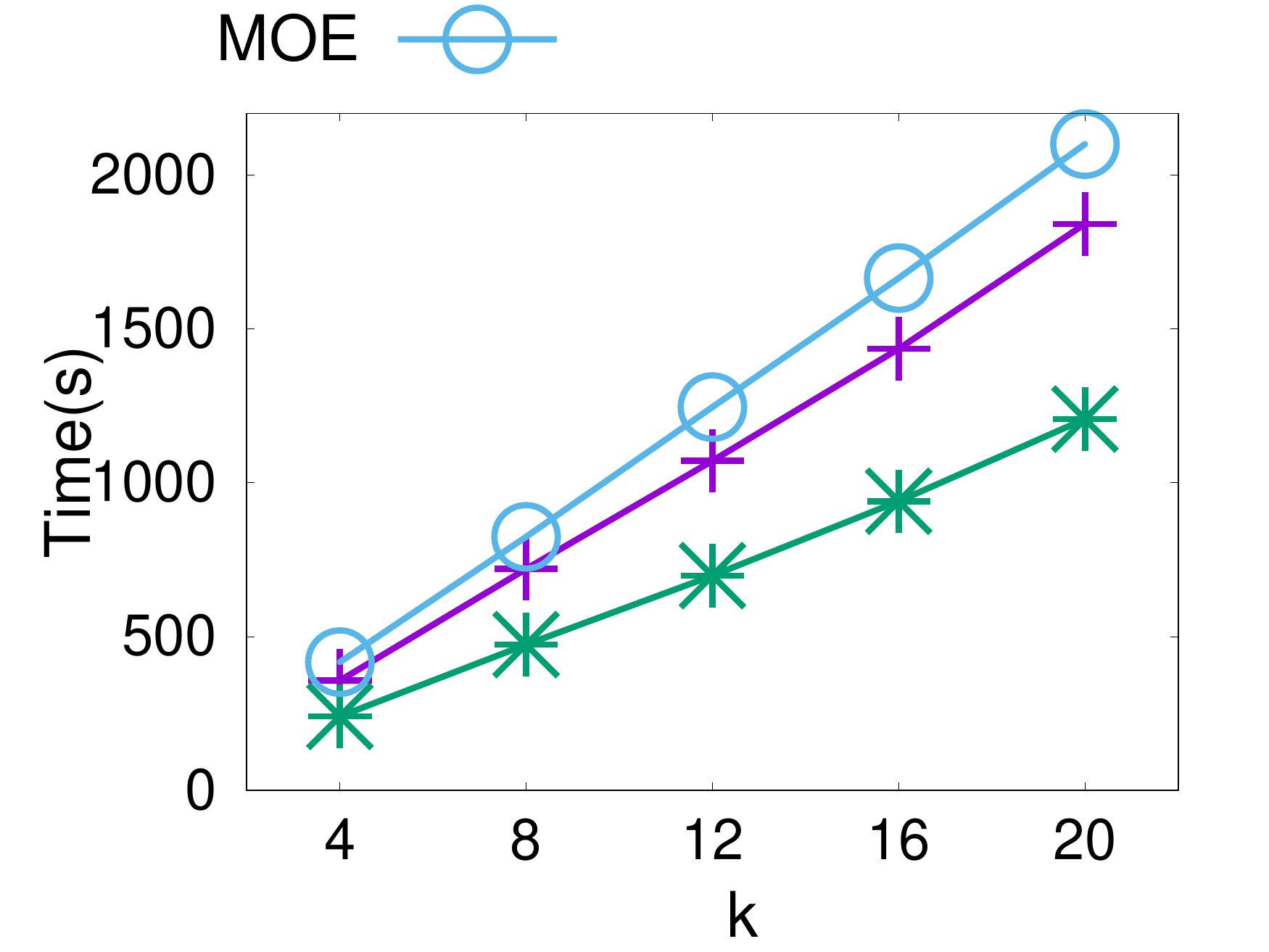}
}
\vspace*{-0.0cm}
\caption{Scalability Test}
\label{fig:exp:scale}
\vspace*{-0.0cm}
\end{figure*}

\begin{figure}[t]
\centering
\subfigure[\GMM]{
\vspace*{-0.4cm}
\label{fig:exp:sssp}
 \includegraphics[width=0.29\columnwidth]{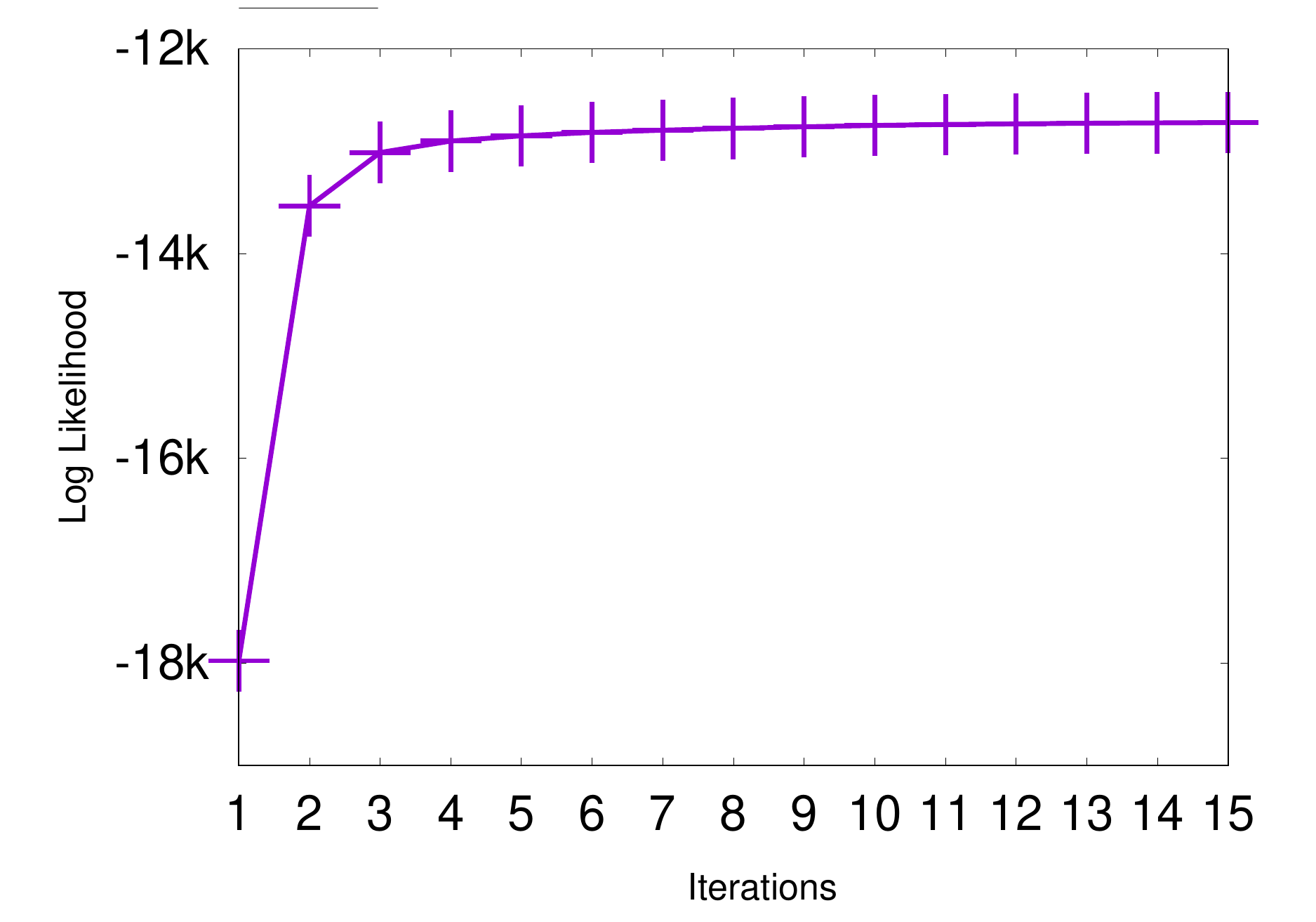}
}
\subfigure[\MLR]{
\label{fig:exp:pagerank}
\vspace*{-0.4cm}
 \includegraphics[width=0.29\columnwidth]{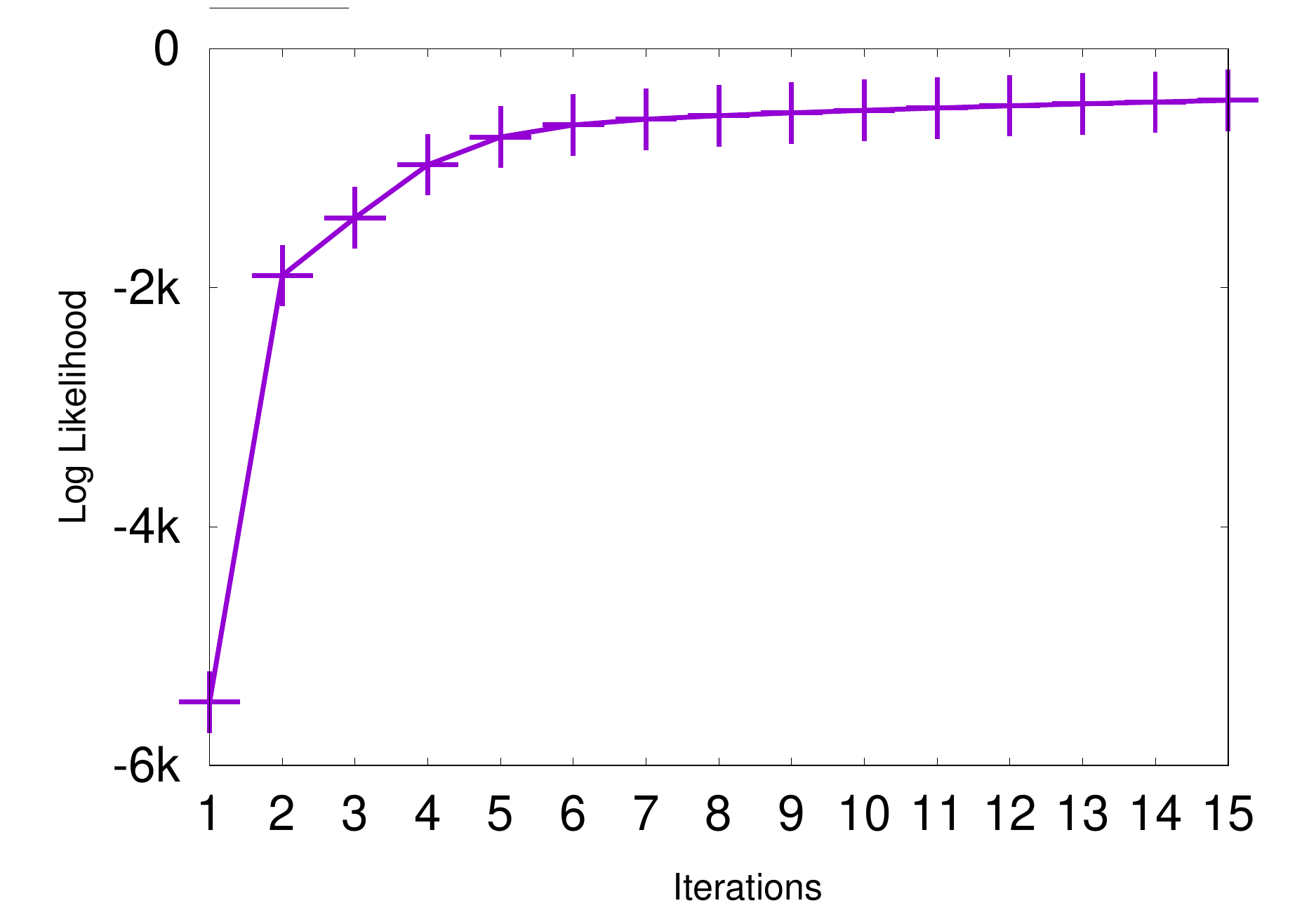}
}
\subfigure[\MOE]{
\vspace*{-0.4cm}
\label{fig:exp:wcc}
 \includegraphics[width=0.29\columnwidth]{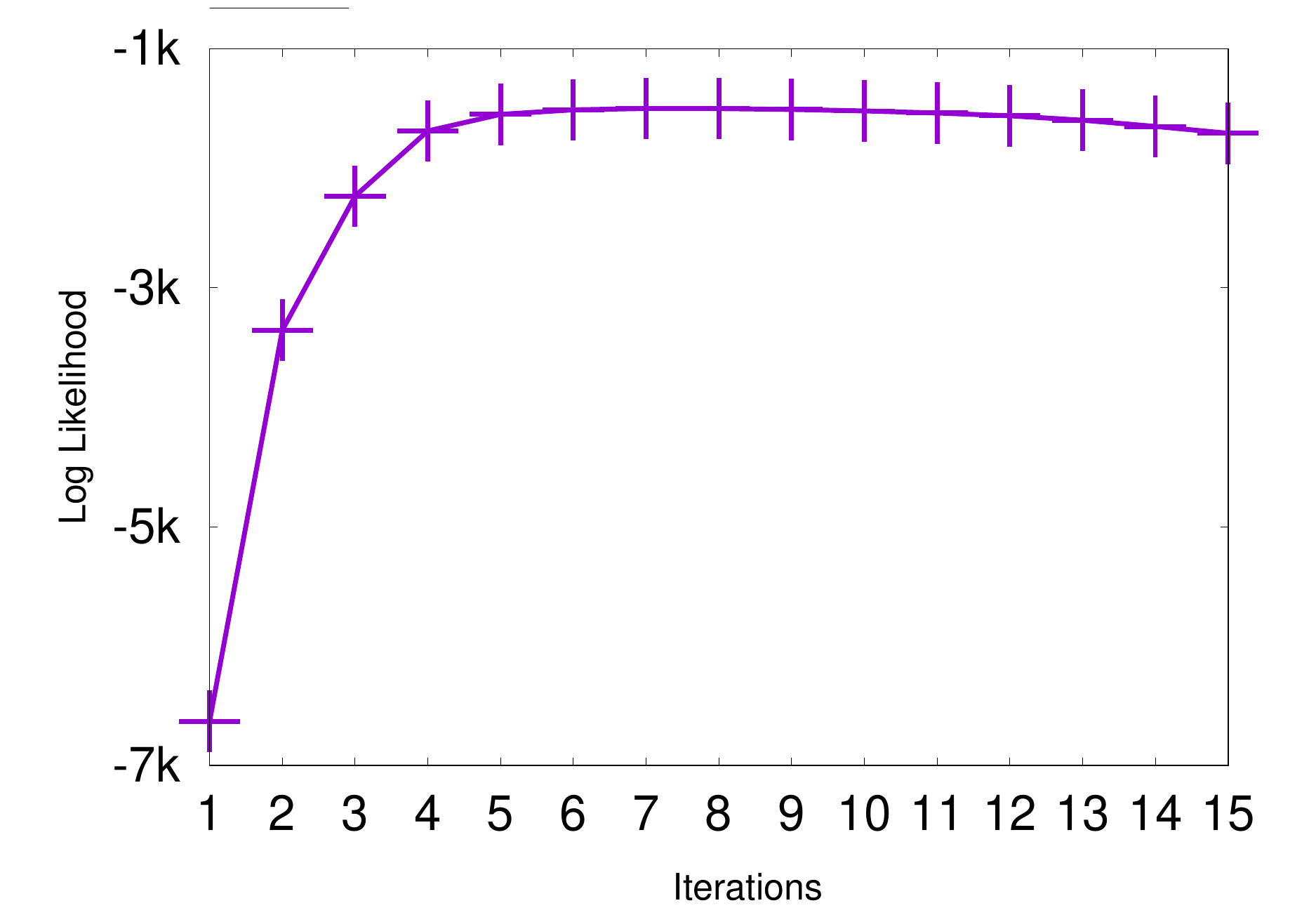}
}
\vspace*{-0.0cm}
\caption{Convergence}
\label{fig:casestudy:converge}
\vspace*{-0.0cm}
\end{figure}

We compare the enhanced \WITH, which translates the recursive \SQL query to \SQL/\PSM with the implementation of using a host language to 
control the looping, which is adopted in pervious EM implementation~\cite{DBLP:conf/sigmod/OrdonezC00}.
We implement the latter by \Psy~\cite{psycopg}, a popular \PostgreSQL adapter for the python language. 
Regarding the EM algorithm, the E-step, M-step, and parameter updating are wrapped in a python for-loop, and executed by a cursor alternatively.
We compare the running time of this two implementations, i.e., enhanced \WITH and \Psy for training Gaussian Mixture Model by varying the dimension $d$ of data point (Fig.~\ref{fig:exp:python:dim}), the scale of the training data $n$ (Fig.~\ref{fig:exp:python:num}), the number of components $k$ (Fig.~\ref{fig:exp:python:k}) and the number of iterations (Fig.~\ref{fig:exp:python:iter}).
The training data is evenly generated from 10 Gaussian distributions.

The evaluated time is the pure query execution time where the costs of database connection, data loading and parameter initialization are excluded.
The experiments show that enhanced \WITH outperforms \Psy significantly, not only for multiple iterators in Fig.~\ref{fig:exp:python:iter} but also for per iteration in Fig.~\ref{fig:exp:python:iter}-~\ref{fig:exp:python:k}. For one thing, the implementation of \Psy calls the databases multiple times per iteration, incurring much client-server communication and context switch costs. 
For the other, the issued queries from client to server will be parsed, optimized and planned on-the-fly. These are the general problems of calling \SQL queries by any host language. 
Meanwhile, we implement the hybrid strategy of \SQLEM~\cite{ordonez2010optimization} on \PostgreSQL. For Gaussian Mixture model, one iteration for 10,000 data points with 10 dimensions fails to terminate within 1 hour. In their implementation, $2k$ separate \SQL queries evaluate the means and variances of 
$k$ components respectively, which is a performance bottleneck. 

\begin{figure*}[t]
\centering
\subfigure[GMM]{
\vspace*{-0.4cm}
\label{fig:casestudy:gmm}
 \includegraphics[width=0.6\columnwidth]{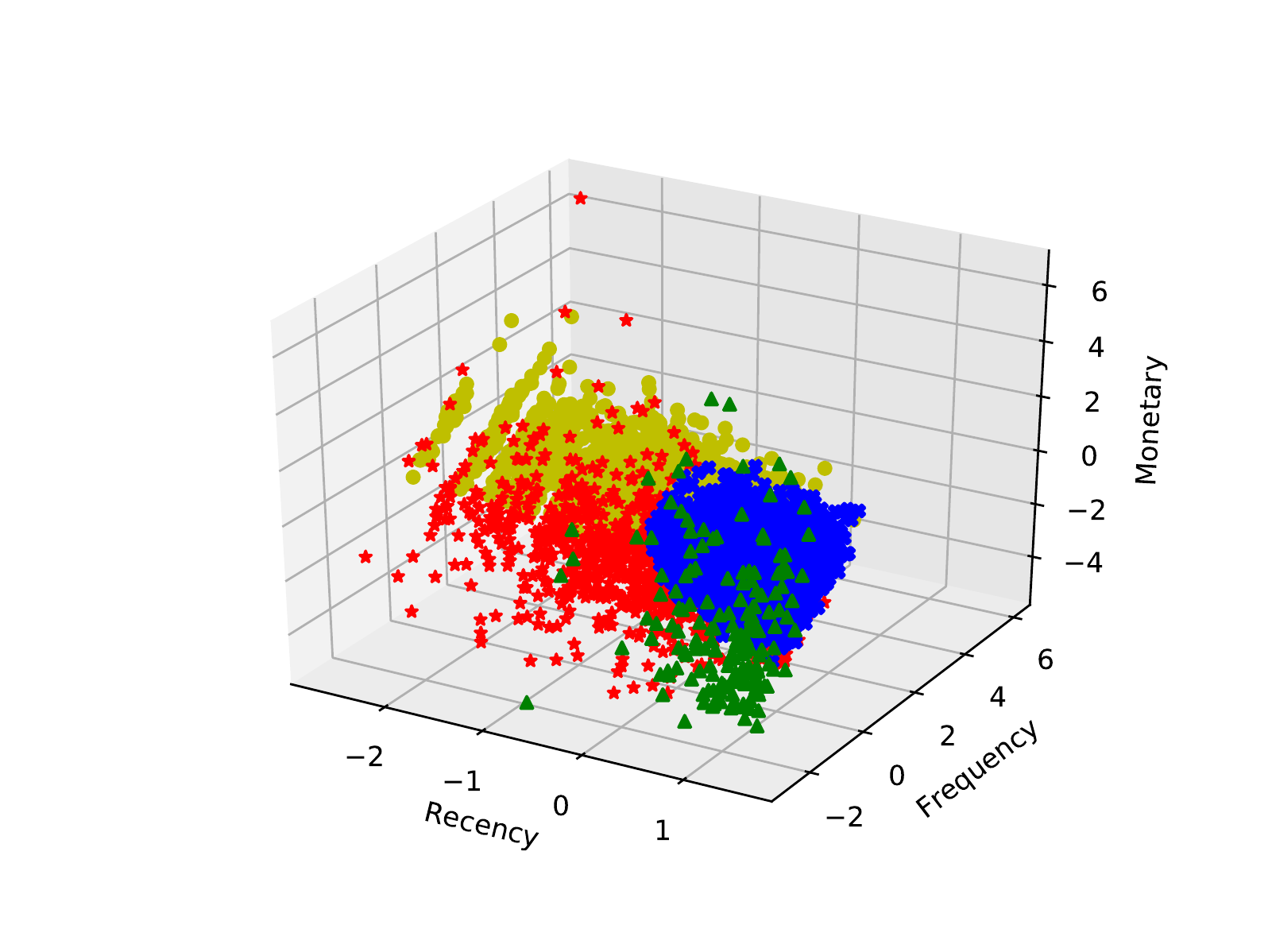}
}
\subfigure[MLR]{
\label{fig:casestudy:lrm}
\vspace*{-0.4cm}
 \includegraphics[width=0.6\columnwidth]{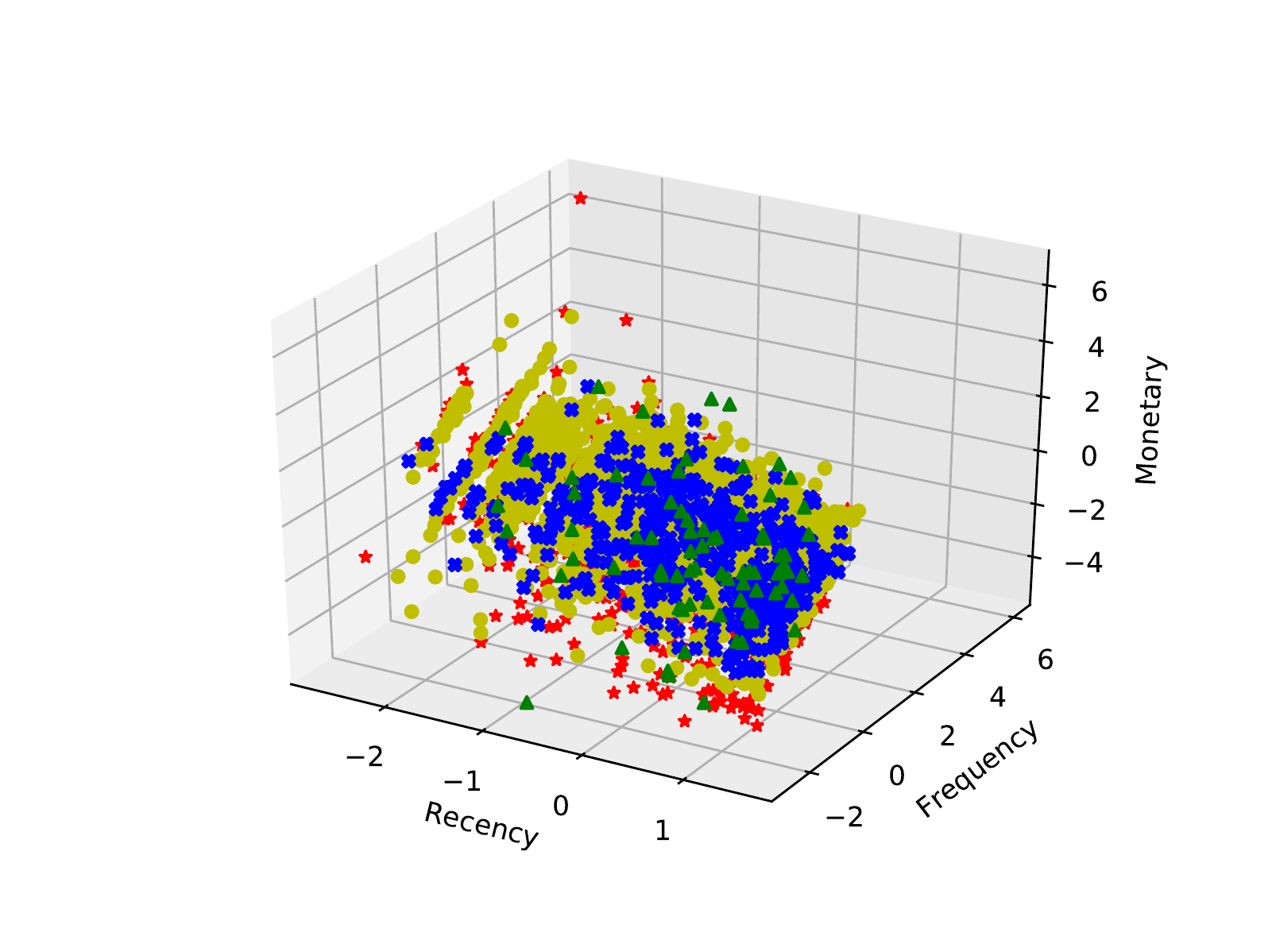}
}
\subfigure[MOE]{
\vspace*{-0.4cm}
\label{fig:casestudy:moe}
 \includegraphics[width=0.6\columnwidth]{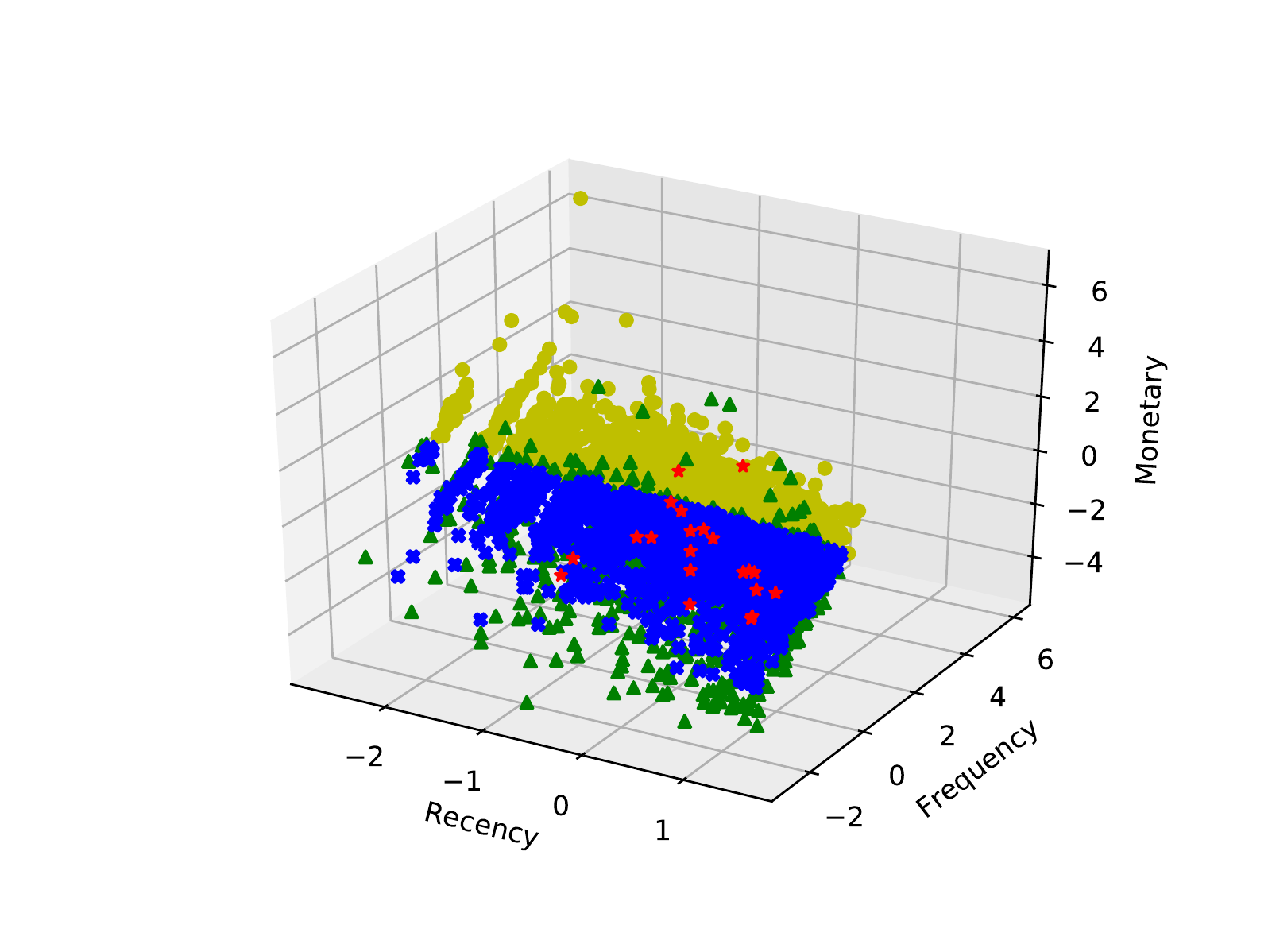}
}
\vspace*{-0.0cm}
\caption{Case Study: Market Segmentation}
\label{fig:exp:casestudy}
\vspace*{-0.0cm}
\end{figure*}

\subsection{Experiments on synthetic data}
We train Gaussian Mixture model (\GMM)~\cite{DBLP:books/lib/Bishop07},  mixture of linear regression (\MLR)~\cite{DBLP:journals/sac/VieleT02} and a neural network model, mixture of experts (\MOE)~\cite{DBLP:journals/tnn/YukselWG12} by evaluating \SQL recursive queries in \PostgreSQL.
Given the observed dataset as $\{(x_1, y1), (x_2, y_1), \cdots, (x_n, y_n)\}$, where $x_{i} \in \mathbb{R}^{d}$ and $y_{i} \in \mathbb{R}$, the \MLR models the density of $y$ given x as 
\begin{equation}
p(y_i|x_{i}) = \sum_{k = 1}^{K} \pi_k \mathcal{N}(y_{i} | x_{i}^{T}\bm{\beta}_{k}, \bm{\sigma}_{k})
\label{eq:lrm}
\end{equation} 
And the \MOE models the density of $y$ given $x$ as
\begin{equation}
p(y_i|x_{i}) = \sum_{k = 1}^{K} g_k(x_{i}) \mathcal{N}(y_{i} | x_{i}^{T}\bm{\beta}_{k}, \bm{\sigma}_{k})
\label{eq:moe}
\end{equation}
where $\bm{\beta}_{k} \in \mathbb{R}^{d}$ is the parameters of a linear transformer, $\mathcal{N}$ is the probability density function of a Gaussian given mean $x_{i}^{T}\bm{\beta}_{k} \in \mathbb{R}$ and standard deviation $\bm{\sigma}_{k} \in \mathbb{R}$.
In Eq.~(\ref{eq:moe}), $g_k(x)$ is called the gating function, given by computing the softmax in Eq.~(\ref{eq:moe:gate}) where $\bm{\theta} \in \mathbb{R}^d$ is a set of linear weights on $x_i$.
\begin{equation}
g_k(x_{i}) = \frac{e^{x_{i}\bm{\theta}_{k}}}{\sum_{j = 1}^{K} e^{x_{i}\bm{\theta}_{j}}}
\label{eq:moe:gate}
\end{equation}
The intuition behind the gating functions is a set of 'soft' learnable weights which determine the mixture of $K$ local models. We adopt the single loop EM algorithm~\cite{DBLP:conf/isnn/YangM09} to estimate the parameters of \MOE, which uses least square regression to compute the gating network directly.  
For \GMM, the training data is evenly drawn from 10 Gaussian distributions.  
For \MLR and \MOE, the training data is generated from 10 linear functions with Gaussian noise. 
The parameters of the Gaussians and the linear functions are drawn from the uniform distribution $[0, 10]$. 
And the initial parameters are also randomly drawn from $[0, 10]$.

Fig.~\ref{fig:exp:scale} displays the training time per iteration of the 3 models by varying the data dimension $d$ (Fig.~\ref{fig:exp:scale:d}), the scale of the training data $n$ (Fig.~\ref{fig:exp:scale:n}) and the number of clusters $k$ (Fig.~\ref{fig:exp:scale:k}).  
In general, for the 3 models, the training time grows linearly as $n$ and $k$ increase, while the increment of data dimension $d$ has a more remarkable impact on the training time.  
When increasing $n$ and $k$, the size of intermediate relations, e.g., relation $R$ for computing the responsibilities in Eq.~(\ref{eq:algebra:estep}) grow linearly. Therefore the training cost grows linearly with regards to $n$ and $k$.
However, in the 3 models, we need to deal with $d \times d$ dimensional matrices in the M-step. 
For \GMM, it needs to compute the probability density of the multivariable Gaussians and reestimate the covariance matrices.
For \MLR and \MOE, they need to compute the matrix inversion and least square regression. 
The training cost grows with regard to the size of matrix. 
The comparison shows it is still hard to scale high-dimensional analysis in a traditional database system. 
However, the efficiency can be improved on a parallel/distributed platform and new hardware.

\begin{figure*}[t]
\centering
\subfigure[n = 100K]{
\vspace*{-0.4cm}
\label{fig:exp:inc:10w}
 \includegraphics[width=0.55\columnwidth]{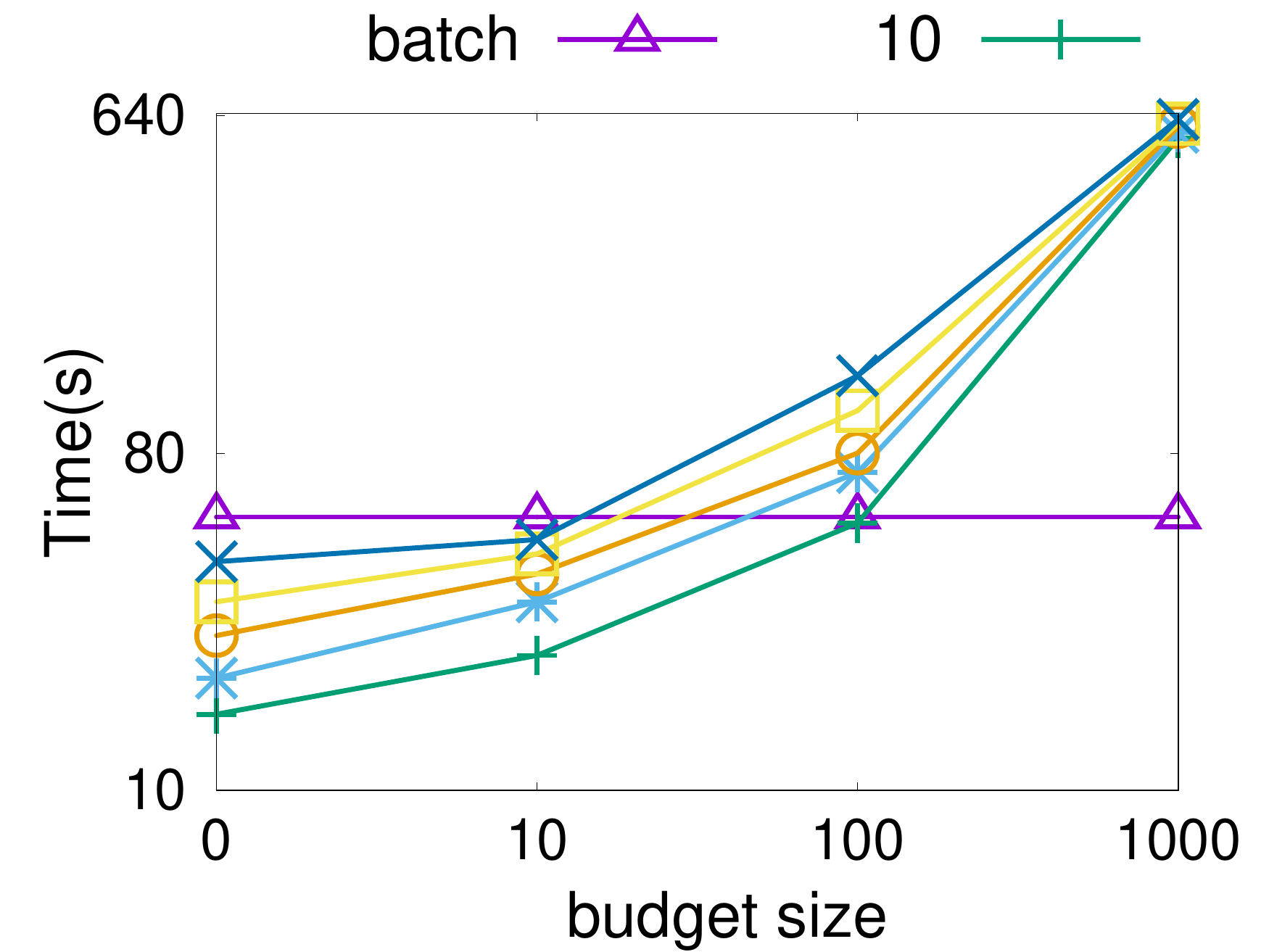}
}
\subfigure[n = 1M]{
\label{fig:exp:inc:100w}
\vspace*{-0.4cm}
 \includegraphics[width=0.55\columnwidth]{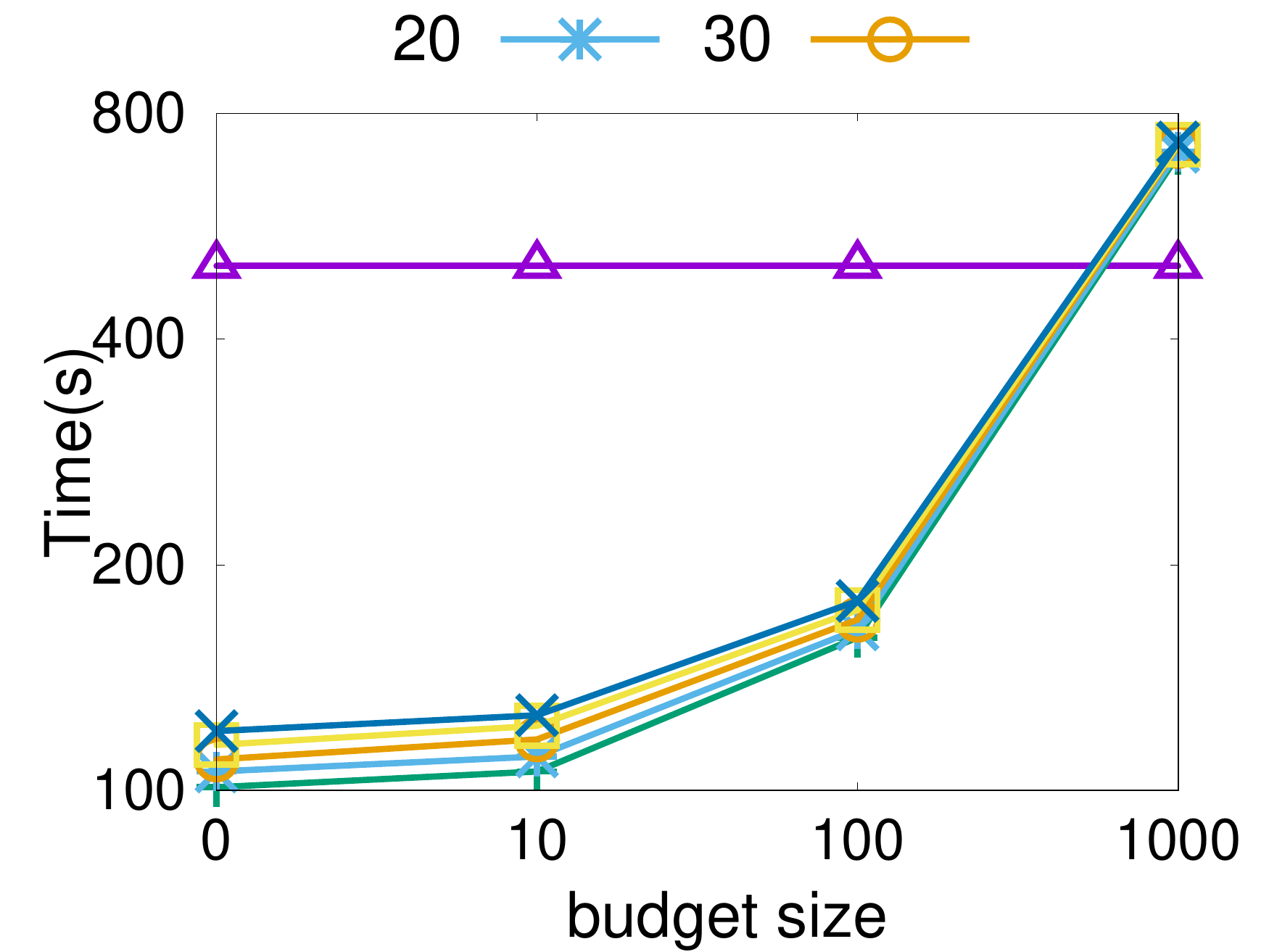}
}
\subfigure[n = 10M]{
\vspace*{-0.4cm}
\label{fig:exp:inc:1000w}
 \includegraphics[width=0.55\columnwidth]{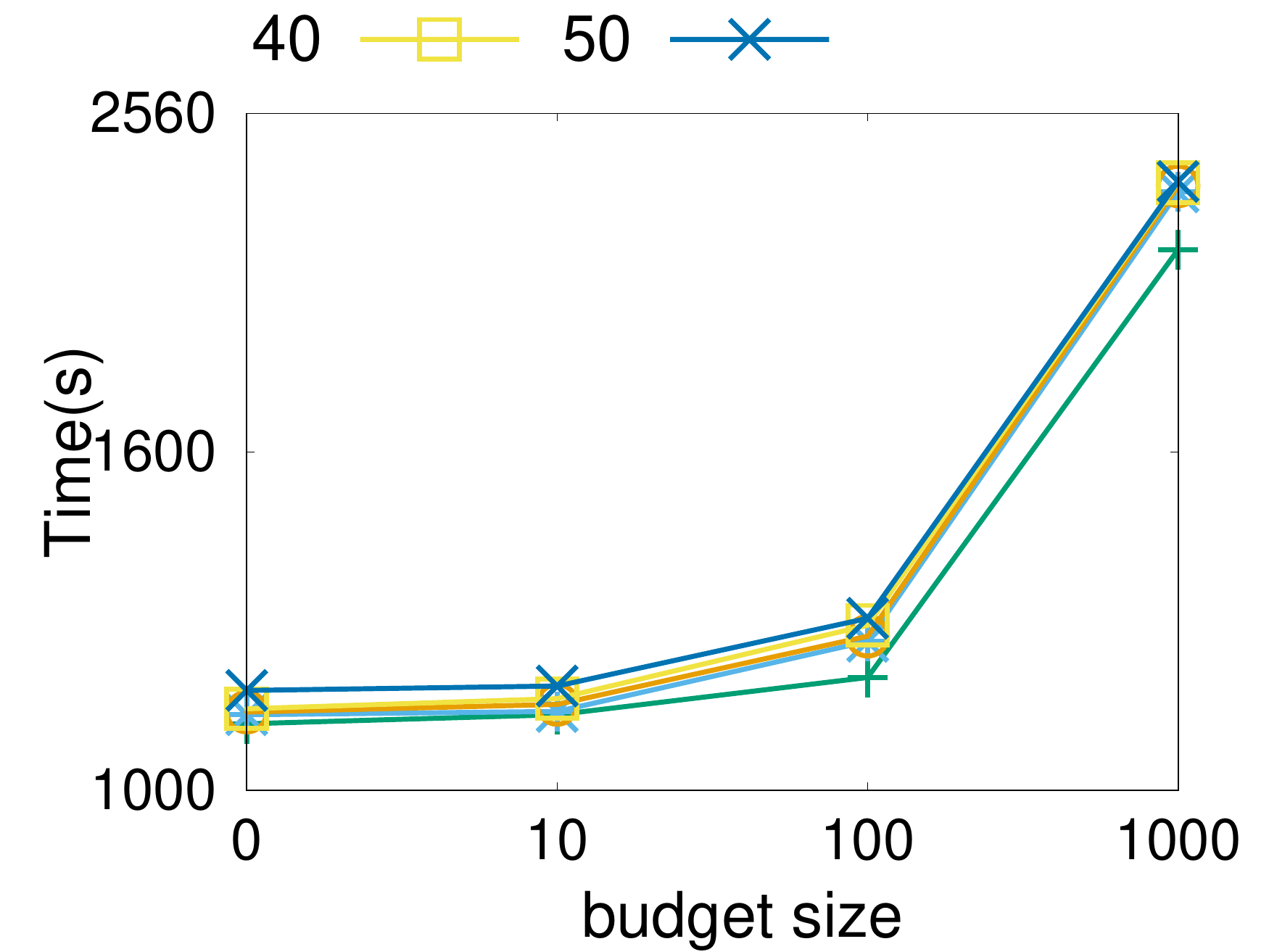}
}
\vspace*{-0.0cm}
\caption{Insert maintenance}
\label{fig:exp:inc}
\vspace*{-0.0cm}
\end{figure*}

\subsection{Case study: market segmentation}
We study building model-based view in \PostgreSQL for a real application, i.e., market segmentation, which partitions the consumers into sub-groups based on their features to analyze their purchase behavior and identify potential market. 
The data is collected from an online retailer, containing 541,908 transactions of 4,308 consumers. 
Based on the RFM model~\cite{chen2012data} for consumer value analysis, for each consumers, a three  dimensional feature of real value including recency, frequency, monetary is extracted. 
The feature is normalized by corresponding means and variances.

We apply the \GMM~(Eq.~(\ref{eq:gmm})), \MLR~(Eq.~(\ref{eq:lrm})) and \MOE~(Eq.~(\ref{eq:moe})) over the  
4,308 consumers and the clustering result is visualized in Fig.~\ref{fig:exp:casestudy}. 
For \MLR and \MOE, the recency and frequency are the features $x$ and monetary is $y$. These two models assume it exists a linear relationship between the recency, frequency and monetary of a consumer. 
All the initial parameters are randomly drawn from the uniform distribution $[0, 1]$.
In Fig.~\ref{fig:exp:casestudy}, the consumers are segmented into 4 clusters by different colors.
The segmentation of \GMM~(Fig.~\ref{fig:casestudy:gmm}) fits data points by Gaussian distributions of 3 variables, while \MLR~(Fig.~\ref{fig:casestudy:lrm}) and \MOE~(Fig.~\ref{fig:casestudy:moe}) fit them by planes with linearity. 
The neural model \MOE tends to generate a sharper cluster boundary than \MLR.
Fig.~\ref{fig:casestudy:converge} shows the convergence of the log-likelihood of the 3 models during 
15 training iterations. In fact, the convergence can be fast and achieved within 5 iterations.

\subsection{Incremental Maintenance}
Finally, we test the performance of our trigger-based model updating mechanism. First, we train \GMM
 for 1-dimensional data generated from 2 Gaussian distributions. The original models are trained over 100k, 1M and 10M data points, respectively with 15 iterations. The overall training time is recorded as the 'batch' mode training time, which is 54s, 501s and 4,841s respectively. 
After the model is trained and persisted. We insert 10, 20, 30, 40, 50 data points to the underlying data by varying the budget size of selected data from 0 to 1,000.

Fig.~\ref{fig:exp:inc} shows the insertion time w.r.t. the budget size of the selected data for the 3 models. The insertion time is the collapsed time from the insert command issuing to the transaction commit, including the cost of data selection with the density-based strategy and computing initial sufficient statistics (line 1 of Algorithm~\ref{algo:modelupdate}). 
As the number of processed tuples increases, the insertion time grows linearly. 
Compare to the retraining cost, i.e., the batch training time, it is not always efficient to update the existing model.
The choice depends on two factors, the size of overall data points, and the budget size plus insert size, i.e., the numbers of data points to be processed in the updating.
The updating mechanism may not be efficient and effective when the overall data size is small or there is a large volume of insertion.  
That is because, for the batch training mode, computation of parameter evaluation dominates the cost. While for the model updating, since the sufficient statistics and the model are updated when processing each data point, the updating overhead becomes a main overhead.
Meanwhile, we notice that the collapsed time of data selection and computing initial sufficient statistics take about 10s, 100s and 1,000s for data size of 100k, 1M and 10M, respectively.  Precomputing and persisting these results will benefit for a larger dataset. 

In this paper, we focus on testing the efficiency of the approximation for model updating. The convergence and approximation guarantee involves a wide range of research topics in statistical and machine learning area, which we leave it as future investigation.  

\section{Conclusion}
\label{sec:conclusion}
Integrating machine learning techniques into database systems facilitates a wide range of applications in industrial and academic fields.  
In this paper, we focus on supporting EM algorithm in \rdbm. Different from the previous approach, our approach wraps the E-step and M-step in an enhanced \SQL recursive query, which is ensured to reach an iterative fix point. The learned model can be materialized as a database view and queries as conventional views. 
Furthermore, to tackle the slight changes of underlying training data, we propose an automatic view updating mechanism by exploiting the incremental variant of the EM algorithm. 
The extensive experiments we conducted validate our enhanced recursive query outperforms previous approach significantly and can support multiple mixture models by EM algorithm, as well as the efficiency of the incremental model update.
It is worth mentioning that the \SQL recursive query is not only suitable for unsupervised learning like EM algorithm, but also has the potential to support supervised learning like classification and regression. 
And the implementation of the query can be migrated to parallel and distributed platforms, e.g., \Hadoop and \Spark, to deploy large scale machine learning applications. 
These directions and convergency and performance guarantee deserve future explorations. 

\section*{Acknowledgement}
This work is supported by the Research Grants Council of
Hong Kong, China under No. 14203618, No. 14202919 and No. 14205520.
{
\bibliographystyle{abbrv}
\pagestyle{plain} 
\bibliography{ref}
}


\end{document}